\documentclass[preprint,secnumarabic,amssymb,amsmath, nofootinbib, aps, prd]{revtex4-1}
\usepackage{bbm} 
\usepackage{graphicx}
\usepackage{hyperref}
\usepackage{bm}
\usepackage{multirow}
\usepackage{color}

\newcommand{\lt}[1]{\tilde{\lambda}_{#1}}
\newcommand{\lh}[1]{\hat{\lambda}_{#1}}
\newcommand{\ft}[1]{\tilde{f}_{#1}}
\newcommand{\dt}[1]{\tilde{d}_{#1}}
\newcommand{\fh}[1]{\hat{f}_{#1}}
\renewcommand{\dh}[1]{\hat{d}_{#1}}
\newcommand{\tr}[1]{\left\langle #1 \right\rangle}
\newcommand{\com}[2]{\left[ #1,#2 \right]}
\newcommand{\acom}[2]{\left\{ #1,#2 \right\}}
\newcommand{\sdot}[2]{\left( #1\cdot #2 \right)}
\newcommand{\Mkmk}{\left( \mu^2_{K^\pm}-\mu^2_{K^0} \right)}

\begin{document}
\title{{\boldmath The axion-baryon coupling in SU(3) heavy baryon chiral perturbation theory}}

\author{Thomas Vonk}
 \email{vonk@hiskp.uni-bonn.de}
 \affiliation{Helmholtz-Institut f\"{u}r Strahlen- und Kernphysik and Bethe Center for Theoretical Physics,
   Universit\"{a}t Bonn, D-53115 Bonn, Germany}
 
 \author{Feng-Kun Guo}
 \email{fkguo@itp.ac.cn}
 \affiliation{CAS Key Laboratory of Theoretical Physics, Institute of Theoretical Physics, Chinese Academy
   of Sciences, Beijing 100190, China}
 \affiliation{School of Physical Sciences, University of Chinese Academy of Sciences,
    Beijing 100049, China}

\author{Ulf-G. Mei{\ss}ner}
 \email{meissner@hiskp.uni-bonn.de}
 \affiliation{Helmholtz-Institut f\"{u}r Strahlen- und Kernphysik and Bethe Center for Theoretical Physics,
   Universit\"{a}t Bonn, D-53115 Bonn, Germany}
 \affiliation{Institute for    Advanced Simulation, Institut f\"ur Kernphysik and J\"ulich Center for Hadron
   Physics,  Forschungszentrum J\"ulich, D-52425 J\"ulich, Germany}
 \affiliation{Tbilisi State University, 0186 Tbilisi, Georgia}

\date{\today}
\begin{abstract}
In the past, the axion-nucleon coupling has been calculated in the framework of SU(2) heavy baryon chiral
perturbation theory up to third order in the chiral power counting. Here, we  extend these earlier
studies to the case of heavy baryon chiral perturbation theory with SU(3) flavor symmetry and derive
the axion coupling to the full SU(3) baryon octet, showing that the axion also significantly couples
to hyperons. As studies on dense nuclear matter suggest the possible existence of hyperons in
stellar objects such as neutron stars, our results should have phenomenological implications
related to the so-called axion window.
\end{abstract}

\maketitle

\section{Introduction}
Soon after the discovery of the instanton solution of euclidean Yang--Mills gauge field theories
\cite{Belavin:1975fg}, it has been realized that the vacuum structure of 
theories such as
quantum chromodynamics (QCD) is highly non-trivial and that a so-called $\theta$-term appears
\cite{Callan:1976je} 
\begin{equation}\label{eq:QCDLagr}
\mathcal{L}_\text{QCD} = \mathcal{L}_\text{QCD,0} + \theta \left(\frac{g}{4\pi}\right)^2
\operatorname{Tr}\left[G_{\mu\nu} \tilde{G}^{\mu\nu}\right],
\end{equation}
which is $CP$ non-invariant as long as no quark is massless and as long as the effective angle
$\bar{\theta}=\theta + \operatorname{Arg}\operatorname{det}\mathcal{M}$ 
is non-zero, where
$\mathcal{M}$ refers to the quark mass matrix. In Eq.~\eqref{eq:QCDLagr}, 
$\mathcal{L}_\text{QCD,0}$
denotes the usual QCD Lagrangian without the $\theta$-term, $g$ is the QCD coupling constant, $G_{\mu\nu}$
the gluon field strength tensor, $\tilde{G}^{\mu\nu}=\epsilon^{\mu\nu\alpha\beta}G_{\alpha\beta}/2$ its dual, and $\operatorname{Tr}$ denotes the
trace in color space. As a consequence, the neutron acquires an electric dipole moment $\propto
\bar{\theta}$~\cite{Baluni:1978rf}. Theoretical estimations of the $\bar{\theta}$-induced
neutron electric dipole moment (nEDM)
roughly vary between $|d_n| \approx 10^{-16} \bar{\theta}\,e\,\text{cm}$ and $|d_n| \approx 10^{-15}
\bar{\theta}\,e\,\text{cm}$; see  Refs.~\cite{Kim:2008hd,Alexandrou:2015spa,Guo:2015tla,Abusaif:2019gry}.
The extreme small upper limit for the physical value of the nEDM determined in experiments
(the most recent result is $|d_n| < 1.8\times 10^{-26}\, e\,\text{cm}$ (90\,\% C.L.)~\cite{Abel:2020gbr}) implies
\begin{equation}
\bar{\theta} \lesssim 10^{-11}~,
\end{equation}
which is a rather unnatural value for a quantity that in principle might take on values between 0
and $2\pi$ (where $\theta=\pi$ is a special point~\cite{Smilga:1998dh}). Even though $\bar{\theta}$
of $\mathcal{O}(1)$ would alter nuclear physics and big bang and stellar nucleosynthesis considerably
\cite{Lee:2020tmi}, such anthropic considerations do not constrain $\bar{\theta}$ to such a
tiny value. One possible way to explain why $\bar{\theta}\approx 0$ (in other words, to resolve
the so-called ``strong $CP$-problem'') is the Peccei--Quinn mechanism~\cite{Peccei:1977hh,Peccei:1977ur},
which features a new global chiral symmetry, now usually labeled $U(1)_\text{PQ}$, and which
automatically leads to $\bar{\theta}=0$. The physical relic of this mechanism is not only
a $CP$-conserving QCD Lagrangian, but also a new, very light pseudoscalar 
pseudo-Nambu--Goldstone
boson with zero bare mass called axion~\cite{Weinberg:1977ma,Wilczek:1977pj}, which soon after
its  theoretical description also became a serious dark matter candidate
\cite{Preskill:1982cy,Abbott:1982af,Dine:1982ah,Ipser:1983mw,Turner:1986tb,Duffy:2009ig,Marsh:2015xka}, possibly
forming Bose--Einstein condensates~\cite{Sikivie:2009qn}. Both, the fact that the Peccei--Quinn
mechanism provides an elegant solution to the strong $CP$-problem, and that it at the same time
might provide a solution to the missing matter problem of the universe, accounts for the unwavering interest
in the axion, even though it has not been detected yet and its existence hence remains hypothetical.

The original (``visible'') Peccei--Quinn--Weinberg--Wilczek (PQWW) axion with a decay constant at the
electroweak scale and a mass in the keV/MeV region seems to be ruled out experimentally
\cite{Donnelly:1978ty,Calaprice:1979pe,Bechis:1979kp,Kim:2008hd}, although there are still attempts
to make the experimental data compatible with the original model~\cite{Alves:2017avw,Liu:2021wap}.
However, these models require a lot of additional assumptions including that these axions restrictively
couple to the first generation fermions (up and down quark, electron) and 
are accidentally pion-phobic.
The clearly preferred models are the so-called ``invisible'' axion models 
such as the
Kim--Shifman--Vainstein--Zakharov (KSVZ) axion model~\cite{Kim:1979if,Shifman:1979if} (which sometimes
is called ``hadronic'' as it does not couple to leptons) or the Dine--Fischler--Srednicki--Zhitnitsky
(DFSZ) axion model~\cite{Dine:1981rt,Zhitnitsky:1980tq}. Phenomenologically, the physical properties
of such an (almost) invisible axion such as its mass and the couplings to 
standard model particles
(quarks, leptons, gauge bosons) are governed by its expectedly large decay constant, which is
traditionally estimated as being~\cite{Preskill:1982cy,Abbott:1982af,Kim:2008hd,Kim:1986ax}
\begin{equation}\label{eq:window}
10^{9}\,\text{GeV} \lesssim f_a \lesssim 10^{12}\,\text{GeV}~,
\end{equation}
such that the axion mass~\cite{Lu:2020rhp,DiLuzio:2020wdo}
\begin{equation}
m_a \approx 5.7 \left(\frac{10^{12}\,\text{GeV}}{f_a} \right) \times 10^{-6}  \,\text{eV}
\end{equation}
presumably would be somewhere between a few $\mu\text{eV}$ and $0.1\,\text{eV}$. The lifetime of
the QCD axion, which may decay into two photons, is incredibly large~\cite{DiLuzio:2020wdo}
\begin{equation}
\tau_a \approx \left(\frac{1\,\text{eV}}{m_a}\right)^5 \times 10^{24}\, \text{s}~.
\end{equation}
Note that the traditional axion window, Eq.~\eqref{eq:window}, is set in order to match cold dark
matter requirements of the canonical invisible axion models that solve the strong $CP$-problem.
However, axions and axion-like particles have been considered also in the 
context of several other
models with decay constants considerably smaller or larger than the window of Eq.~\eqref{eq:window}.
For example, in the case of hadronic axions a decay constant as small as $f_a\approx 10^{6}\,\text{GeV}$
has been proposed~\cite{Chang:1993gm}, while in string theories axions and axion-like particles
might appear at the GUT or even the Planck scale, i.e. $10^{15}\,\text{GeV} \lesssim f_a \lesssim 10^{18}\,
\text{GeV}$~\cite{Svrcek:2006yi}.

As a consequence of its (model-dependent) coupling to standard model particles, the axion of
course also couples to composite particles such as mesons or baryons. One 
possible method of
estimating constraints on the axion decay constant (or equivalently its mass) rests upon this coupling
to baryons, in particular to nucleons, namely through nuclear bremsstrahlung processes in stellar objects
\cite{Iwamoto:1984ir,Mayle:1987as,Brinkmann:1988vi,Raffelt:1987yt,Turner:1987by,Burrows:1988ah,Keil:1996ju,Raffelt:2006cw,Keller:2012yr,Sedrakian:2015krq,Hamaguchi:2018oqw,Beznogov:2018fda,Sedrakian:2018kdm,Chang:2018rso,Carenza:2019pxu} (see also the overviews~\cite{Raffelt:1990yz,Turner:1989vc,Kim:2008hd,DiLuzio:2020wdo}).  The leading order $aN$ coupling has been derived several times since the early days
\cite{Donnelly:1978ty,Kaplan:1985dv,Srednicki:1985xd,Georgi:1986df,Chang:1993gm}. More recently,
this leading order coupling has been reexamined in Ref.~\cite{diCortona:2015ldu}. Moreover, such an analysis 
has been carried out up-to-and-including  $\mathcal{O}(p^3)$ in SU(2) heavy baryon chiral perturbation
theory (HBCHPT) in our previous work~\cite{Vonk:2020zfh}, where $p$
denotes a small parameter (see below in section~\ref{ch:HBCHPT}).

In this paper, we  extend these previous studies using SU(3) HBCHPT
in two regards: (i) we extend the calculations of the axion-nucleon couplings to the SU(3) case,
up-to-and-including $\mathcal{O}(p^3)$, and (ii) we derive the couplings of the axion to the full
ground state baryon octet (section~\ref{ch:results}). At this point one might
wonder whether this extension is indeed useful given the fact that the most relevant coupling of
the axion to baryons, i.e. the coupling to nucleons, is already known to good precision in the
literature cited above, certainly precise enough for any phenomenological purpose. In this sense, the
present study on the one hand is of rather theoretical interest, unraveling the explicit
NNLO structure of the axion-nucleon coupling explicitly respecting the strange quark mass dependence.
On the other hand, however, extending the analysis to the SU(3) case, we show that the traditional
``invisible'' axion also significantly couples to hyperons which has further phenomenological
implications, since it has been suggested that hyperons might exist in the cores of neutron stars \cite{Glendenning:1982nc,Glendenning:1984jr,Maxwell:1986pj,Weber:1989hr,Ellis:1990qq,Glendenning:1991es,Knorren:1995ds,Schaffner:1995th,Balberg:1998ug,Baldo:1999rq,Shen:2002qg,Lackey:2005tk,Djapo:2008au,Bednarek:2011gd,Weissenborn:2011ut,Miyatsu:2013yta,Lopes:2013cpa,Fortin:2014mya,Oertel:2014qza,Katayama:2015dga,Chatterjee:2015pua,Tolos:2016hhl,Tolos:2017lgv,Negreiros:2018cho,Sun:2018tmw,Ofengeim:2019fjy,Fortin:2020qin,Tolos:2020aln}.
While it is a matter of ongoing research whether the seemingly energetically inevitable existence of
hyperons in dense cores are compatible with the observed maximum masses of neutron stars (this question
is related to the expected softening of the equation of state and known in the literature as
``hyperon puzzle''), it is clear that approaches to constrain axion properties from cooling of
neutron stars based on axion-nucleon bremsstrahlung alone might turn out to be insufficient.
Depending on the underlying model and parameter sets, particularly the $\Lambda$ and the
$\Sigma^-$ might appear in significant fractions of the total baryon number. As will be shown
in this study, the axion coupling to these particular hyperons is of a similar order as
that to the nucleons (in fact, it could even be much larger than that to the neutron depending on the axion models), suggesting a revision of axion parameter constraints from stellar cooling.

Before performing the calculations of the axion-baryon couplings, we start with the explicit
implementation of our framework and work out its ingredients and building 
blocks, which are the main
topics of the following section~\ref{ch:model}. This section ends with some remarks of the general form 
of the axion-baryon coupling (section~\ref{ch:generalshape}). The actual calculations of the axion-baryon
couplings up to the leading one-loop level are performed in section~\ref{ch:determiningcoupl}, and the results
are discussed in section~\ref{ch:results}.

\section{SU(3) chiral perturbation theory with axions: General remarks}\label{ch:model}

\subsection{Ingredients from the axion-quark interaction Lagrangian}

Consider the general QCD Lagrangian with axions below the electroweak symmetry breaking scale~\cite{Kim:1986ax}
\begin{equation}\label{eq:a-q-lagrangian0}
\mathcal{L}_{QCD} = \mathcal{L}_\text{QCD,0} + \frac{a}{f_a} \left(\frac{g}{4\pi}\right)^2
\operatorname{Tr}\left[G_{\mu\nu} \tilde{G}^{\mu\nu}\right] + \bar{q}\gamma^\mu  \gamma_5
\frac{\partial_\mu a}{2f_a} \mathcal{X}_q q~,
\end{equation}
where $q=(u,d,s,c,b,t)^\mathrm{T}$ collects the quark fields and $a$ refers to the axion field
with decay constant $f_a$. The second term is a remnant of the $\theta$-term of Eq.~\eqref{eq:QCDLagr}
after the spontaneous breakdown of the Peccei--Quinn symmetry. Depending on the underlying axion model,
axions might additionally couple directly to the quark fields, which is here present in the form of the
last term. Here, we assume the canonical scenario that the couplings are flavor conserving at tree-level,
i.e $\mathcal{X}_q=\operatorname{diag}\left\{X_q\right\}$ is a diagonal 
$6\times 6$ matrix acting in
flavor space, where the $X_q$'s, $q=\left\{u,d,s,c,b,t\right\}$, are the coupling constants of the
respective axion-quark interactions. These are given, for instance, by 
\begin{align}
\label{eq:couplignconstantsmodeldepending}
\begin{split}
X_q^\mathrm{KSVZ} & = 0 \ , \\
X_{u,c,t}^\mathrm{DFSZ} & = \dfrac{1}{3} \dfrac{x^{-1}}{x+x^{-1}} = \dfrac{1}{3}\sin^2\beta\ , \\
X_{d,s,b}^\mathrm{DFSZ} &=\dfrac{1}{3} \dfrac{x}{x+x^{-1}}=\dfrac{1}{3}\cos^2\beta = \dfrac{1}{3}
- X_{u,c,t}^\mathrm{DFSZ} \ , 
\end{split}
\end{align}
for the KSVZ-type axions and DFSZ-type axions, respectively, where $x=\cot\beta$ is the ratio of the
vacuum expectation values (VEVs) of the two Higgs doublets within the latter models.
Note that $\mathcal{L}_\text{QCD,0}$ in Eq.~\eqref{eq:a-q-lagrangian0} also contains a quark mass
term $\bar{q}\mathcal{M}_q q$, with $\mathcal{M}_q = \operatorname{diag}\left\{m_q\right\}$ being the real,
diagonal and $\gamma_5$-free quark mass matrix.

As usual, it is advisable to perform a transformation on the quark fields,
\begin{equation}
q \to \exp\left({i\gamma_5\frac{a}{2f_a}\mathcal{Q}_a}\right)\,q~,
\end{equation}
in order to remove the second term of Eq.~\eqref{eq:a-q-lagrangian0}. Choosing
\begin{equation}
  \mathcal{Q}_a=\frac{\mathcal{M}_q^{-1}}{\tr{\mathcal{M}_q^{-1}}}\approx
  \frac{1}{1+z+w}\operatorname{diag}\left(1,z,w,0,0,0\right) ,
\end{equation} 
which corresponds to vacuum alignment in the $\theta$-vacuum case, we can 
avoid the leading
order mass mixing between the axion and the neutral pseudo-Nambu--Goldstone bosons of the spontaneous breaking of SU(3) chiral symmetry, the $\pi^0$ and the $\eta$, from the beginning. Here, $\tr{\dots}$ denotes
the trace in flavor space, $z=m_u/m_d$ and $w=m_u/m_s$.
Now the axion-quark interaction Lagrangian is given by
\begin{equation}\label{eq:a-q-lagrangian1}
  \mathcal{L}_{a\text{--}q} = - \left(\bar{q}_L\mathcal{M}_a q_R + \text{h.c.}\right) + \bar{q}\gamma^\mu
  \gamma_5 \frac{\partial_\mu a}{2f_a} \left(\mathcal{X}_q- \mathcal{Q}_a 
\right) q~,
\end{equation}
where $q_L$ and $q_R$ are the left- and right-handed projections of the quark fields, and 
\begin{equation}\label{eq:Mass:a}
\mathcal{M}_a =\exp\left({i\frac{a}{f_a}\mathcal{Q}_a}\right) \, \mathcal{M}_q~.
\end{equation}
Considering only the three-dimensional subspace of flavor space, i.\,e. $q=\left(u,d,s\right)^\text{T}$,
and introducing
\begin{align}
c^{(1)} &= \frac{1}{3}\left(X_u+X_d+X_s-1\right) ,\nonumber\\
c^{(3)} &= \frac{1}{2}\left(X_u-X_d-\frac{1-z}{1+z+w}\right) ,\\
c^{(8)} &= \frac{1}{2\sqrt{3}}\left(X_u+X_d-2X_s-\frac{1+z-2w}{1+z+w}\right) ,\nonumber
\end{align}
we can decompose the matrix $\mathcal{X}_q- \mathcal{Q}_a$ into traceless 
parts and parts with
non-vanishing trace, so that
\begin{align}\label{eq:a-q-lagrangian2}
\begin{split}
  \mathcal{L}_{a\text{--}q} = & - \left(\bar{q}_L\mathcal{M}_a q_R + \text{h.c.}\right)\\ & + \left(\bar{q}\gamma^\mu  \gamma_5 \frac{\partial_\mu a}{2f_a} \left(c^{(1)}\mathbbm{1}+c^{(3)} \lambda_3+c^{(8)}\lambda_8\right) q\right)_{q=(u,d,s)^\text{T}}\\ & +\sum_{q=\{c,b,t\}}\left(\bar{q}\gamma^\mu  \gamma_5 \frac{\partial_\mu a}{2f_a} X_q q\right) ,
\end{split}
\end{align}
where $\lambda_3$ and $\lambda_8$ refer to the third and eighth Gell-Mann 
matrices, respectively.
From this form of the axion-quark interaction Lagrangian, we can directly read off the required
ingredients for the axionic SU(3) heavy baryon chiral Lagrangian, i.\,e. the external currents
\begin{align}\label{eq:externalcurrents}
\begin{split}
  s & = \mathcal{M}_a~,\\
  a_\mu & = \frac{\partial_\mu a}{2f_a} \left(c^{(3)} \lambda_3+c^{(8)}\lambda_8\right) ,\\
    a^{(s)}_{\mu,i} & = c_i \frac{\partial_\mu a}{2f_a} \mathbbm{1}~,\qquad\qquad\qquad i=1,\dots, 4~.
\end{split}
\end{align}
Here we have set
\begin{equation}
c_1=c^{(1)}~,\qquad c_2=X_c~,\qquad c_3=X_b~,\qquad c_4=X_t~. 
\end{equation}
Note that in contrast to usual chiral perturbation theory, it is also necessary to add isosinglet
axial-vector currents  $a_{\mu,i}^{(s)}$ in order to preserve the full QCD axion interaction. This
is possible here, as the subtleties that usually arise due to the U(1)$_A$ anomaly are absent, because
the model is now anomaly-free.

If one considers also flavor-changing axion-quark couplings at tree-level, $\mathcal{X}_q$ would
be non-diagonal. In this case, it is likewise appropriate to decompose $\mathcal{X}_q- \mathcal{Q}_a$
into traceless parts and parts with non-vanishing trace, such that 
\begin{equation}\label{eq:externalcurrents2}
  a_\mu = \frac{\partial_\mu a}{2f_a} \sum_{i=1}^8 C^{(i)} \lambda_i~,
\end{equation}
with $C^{(i)}$ depending on the allowed flavor-changing processes and $\lambda_i$ the eight Gell-Mann matrices. This then would cause baryon
conversion processes. At this point we note that depending on the underlying model Higgs loops may induce off-diagonal couplings, which would be loop-suppressed but might be relevant due to unsuppressed top Yukawa couplings  \cite{Choi:2021kuy}. As stated above, we here stick to the most prevalent models excluding such
flavor-changing axion-quark couplings at tree-level.

\subsection{Building blocks of the chiral Lagrangian}\label{ch:HBCHPT}

The framework we use in this paper is chiral perturbation theory (CHPT) including baryons, meaning
the effective field theory of QCD in the low-energy sector. CHPT is used to explore  meson-baryon
systems based on a systematic expansion in small momenta and quark masses 
below the scale of the
spontaneous breakdown of chiral symmetry, $\Lambda_\chi\sim 1$\,GeV, as first developed by Gasser, Sainio,
and \v{S}varc for the two-flavor case~\cite{Gasser:1987rb} and by Krause for the three-flavor
case~\cite{Krause:1990xc}. However, while baryons can easily be incorporated into CHPT in a  consistent manner respecting all symmetries, the power-counting scheme engineered to
systematically arrange the infinitely many terms allowed by the fundamental symmetries is spoiled
by the fact that the baryon masses $m_B$, which do not vanish in the chiral limit, are roughly of the
same order as $\Lambda_\chi$. One way to overcome this problem is to realize that all mass scales need to be assigned to a scaling in the power counting and thus to treat the baryons as extremely
heavy, static fermions, which is the idea behind HBCHPT~\cite{Jenkins:1990jv, Bernard:1992qa, Bernard:1995dp, Muller:1996vy}. In this scheme, the four momentum of a baryon field is decomposed as
\begin{equation}
k_\mu = m_B v_\mu + p_\mu
\end{equation}
with the four-velocity $v_\mu$ subject to the constraint $v^2=1$, and $p_\mu$ a small residual
momentum satisfying $\sdot{v}{p} \ll m_B$. After integrating out the heavy degrees of freedom,
the resulting Lagrangian is expressed in terms of the heavy baryon field $B$, which is now characterized
by a fixed velocity $v_\mu$ (for convenience, we refrain from writing $B_v$ to explicitly mark the
$v$ dependence). Additionally, it is possible to simplify the Dirac algebra by expressing any
Dirac bilinear by means of $v_\mu$ and the Pauli--Lubanski spin operator
\begin{equation}
S_\mu=\frac{i}{2} \gamma_5 \sigma_{\mu\nu} v^\nu~,
\end{equation}
which satisfies the relations
\begin{equation}\label{eq:spinoperator}
\acom{S_\mu}{S_\nu} = \frac{1}{2} (v_\mu v_\nu - g_{\mu\nu})~,\quad\com{S_\mu}{S_\nu} =
i \epsilon_{\mu\nu\rho\sigma} v^\rho S^\sigma~,\quad \sdot{v}{S} = 0~,\quad S^2 = \frac{1-d}{4},
\end{equation}
where the last one is valid in $d$ spacetime dimensions. In Eq.~\eqref{eq:spinoperator} and in what follows,
$\com{\ }{\ }$ refers to the commutator and $\acom{\ }{\ }$ to the anticommutator.

Any term in the effective Lagrangian consists of a number of elements from a small set of
basic building blocks, from which we will present those that are relevant 
for axionic
SU(3) HBCHPT. In this three-flavor case, the ground state baryon octet consisting of the
nucleons $p$ and $n$ and the hyperons $\Sigma$, $\Lambda$, and $\Xi$ are collected in a single
$3\times 3$ matrix
\begin{equation}\label{eq:baryons}
B = \begin{pmatrix}
\frac{1}{\sqrt{2}} \Sigma_3  + \frac{1}{\sqrt{6}} \Lambda_8 & \Sigma^+ & p \\
\Sigma^- & -\frac{1}{\sqrt{2}} \Sigma_3 + \frac{1}{\sqrt{6}} \Lambda_8 & n \\
\Xi^- & \Xi^0 & -\frac{2}{\sqrt{6}} \Lambda_8 
\end{pmatrix},
\end{equation}
where
\begin{align}
\begin{split}
\Sigma_3 & = \cos\epsilon\ \Sigma^0 -\sin\epsilon\ \Lambda~, \\
\Lambda_8 & = \sin\epsilon\ \Sigma^0 +\cos\epsilon\ \Lambda~.
\end{split} 
\end{align}
Equation~\eqref{eq:baryons} corresponds to the adjoint representation of SU(3). Due to isospin breaking,
the physical $\Sigma^0$ and $\Lambda$ are mixed states made of the $\Sigma_3$ and $\Lambda_8$.
This mixing is usually parameterized by the mixing angle $\epsilon$ and one has
\begin{equation}\label{eq:epsmixing}
\tan{2\epsilon} = \frac{\tr{\lambda_3 \mathcal{M}_q}}{\tr{\lambda_8 \mathcal{M}_q}}~.
\end{equation}
The pseudoscalar mesons, the pseudo-Nambu--Goldstone bosons of the spontaneous breakdown of chiral symmetry,
appear in the Lagrangian in form of a unitary matrix
\begin{equation}
u=\sqrt{U}=\exp\left(i\frac{\Phi}{2F_p}\right),
\end{equation}
where $F_p$ is the pseudoscalar decay constant in the chiral limit and $\Phi$ is a Hermitian $3\times 3$
matrix given by
\begin{equation}\label{eq:mesons}
\Phi = \sqrt{2} \begin{pmatrix}
\frac{1}{\sqrt{2}} \pi_3 + \frac{1}{\sqrt{6}} \eta_8 & \pi^+ & K^+ \\
\pi^- & -\frac{1}{\sqrt{2}} \pi_3 + \frac{1}{\sqrt{6}} \eta_8 & K^0 \\
K^- & \bar{K^0} & -\frac{2}{\sqrt{6}} \eta_8
\end{pmatrix}.
\end{equation}
Again, the physical mass eigenstates of the neutral particles of the diagonal are mixed states as
a consequence of isospin breaking effects and one has
\begin{align}
\begin{split}
\pi_3 & = \cos\epsilon\ \pi^0 -\sin\epsilon\ \eta~, \\
\eta_8 & = \sin\epsilon\ \pi^0 +\cos\epsilon\ \eta~.
\end{split} 
\end{align}
Using the leading order meson masses,
\begin{align}\label{eq:mesonmasses}
\begin{split}
M_{\pi^\pm}^2 & = B_0(m_u+m_d)~,\\
M_{\pi^0}^2 & = B_0(m_u+m_d)+\frac{2}{3}B_0(m_u+m_d-2m_s)\frac{\sin^2\epsilon}{\cos 2\epsilon}~,\\
M_{K^\pm}^2 & = B_0(m_u+m_s)~,\\
M_{K^0}^2 & = B_0(m_d+m_s)~,\\
M_{\eta}^2 & = \frac{1}{3}B_0(m_u+m_d+4m_s)-\frac{2}{3}B_0(m_u+m_d-2m_s)\frac{\sin^2\epsilon}{\cos 2\epsilon}~,
\end{split}
\end{align}
where $B_0$ is a parameter from the $\mathcal{O}(p^2)$ meson Lagrangian related to the
scalar quark condensate. One can further derive 
\begin{align}\label{eq:mixingmesonmasses}
\begin{split}
\sin 2\epsilon & = \frac{2}{\sqrt{3}}\frac{M_{K^\pm}^2-M_{K^0}^2}{M_{\pi^0}^2-M_{\eta}^2}~,\\
\cos 2\epsilon & = \frac{2}{3}\frac{2M_{\pi^\pm}^2-M_{K^\pm}^2-M_{K^0}^2}{M_{\pi^0}^2-M_{\eta}^2}~,
\end{split}
\end{align}
using Eq.~\eqref{eq:epsmixing}, which is valid at leading order.

Any other particle such as the axion enters the theory in form of an external current. In the present
case, we need five axial-vector currents $a_\mu$ and $a_{\mu,i}^{(s)}$, $i=1,\dots,4$ corresponding to the isovector
and isoscalar parts of the axion-matter interaction; compare Eq.~\eqref{eq:externalcurrents}. Furthermore, a scalar external field $s$ is
needed to account for the explicit chiral symmetry breaking due to the light quark masses. Setting $\chi=2B_0 s$,
it is then convenient to define
\begin{align}\label{eq:chipm}
\begin{split}
   u_\mu & = i\left[ u^\dagger \partial_\mu u - u \partial_\mu u^\dagger - i u^\dagger
  a_\mu u - i u a_\mu u^\dagger \right],\\ 
  u_{\mu, i} & = i\left[ - i u^\dagger
  a^{(s)}_{\mu, i} u - i u a^{(s)}_{\mu, i} u^\dagger \right] = 2a^{(s)}_{\mu, i}~,\\
  \chi_\pm & = u^\dagger\chi u^\dagger \pm u\chi^\dagger u~,
\end{split}
\end{align}
which all transform the same way under chiral transformations. Additionally, one defines the
chiral covariant derivative
\begin{equation}
\com{\mathcal{D}_\mu}{B} = \partial_\mu B + \com{\Gamma_\mu}{B},
\end{equation}
where
\begin{equation}
   \Gamma_\mu  = \frac{1}{2}\left[ u^\dagger \partial_\mu u + u \partial_\mu u^\dagger - i u^\dagger
  a_\mu u + i u a_\mu u^\dagger \right]
\end{equation}
is the chiral connection. Assigning systematically a chiral dimension $p$ 
to these building blocks
in order to establish the power-counting method mentioned above, the effective Lagrangian is constructed
by considering all combinations allowed by the underlying symmetries. The 
terms of the meson-baryon
Lagrangian then can be arranged according to the chiral dimension,
\begin{equation}\label{eq:lagrpower}
  \mathcal{L}_{\Phi B} = \mathcal{L}_{\Phi B}^{(1)} + \mathcal{L}_{\Phi 
B}^{(2)} + \mathcal{L}_{\Phi B}^{(3)}
  + \dots + \mathcal{L}_{\Phi}^{(2)}+ \mathcal{L}_{\Phi}^{(4)} + \dots~,
\end{equation}
where the superscript ``${(i)}$'' denotes the Lagrangian containing all terms of $\mathcal{O}(p^i)$.
In the case of the purely mesonic Lagrangians $\mathcal{L}^{(i)}_{\Phi}$, 
$i$ is restricted to even numbers.
Chiral loops start contributing at $\mathcal{O}(p^3)$ (see section~\ref{ch:loops}), which means that
$\mathcal{L}_{\Phi B}^{(3)}$ contains a number of low-energy constants and counter-terms needed
for the renormalization (ch.~\ref{ch:NNLOchip} and ~\ref{ch:NNLOct}). Note that in non-relativistic
HBCHPT the Lagrangian~\eqref{eq:lagrpower} contains an inverse power series in $m_B$, the average
octet mass in the chiral limit, which starts contributing at $\mathcal{O}(p^2)$ (see section~\ref{ch:mBexpansion}).
The leading order meson-baryon Lagrangian is given by 
\begin{equation}\label{eq:LagrLO}
  \mathcal{L}_{\Phi B}^{(1)} = \tr{ i \bar{B} v^\mu \com{\mathcal{D}_\mu}{B}} + D \tr{\bar{B} S^\mu \acom{u_\mu}{B}}
  + F\tr{\bar{B} S^\mu \com{u_\mu}{B}} + D^i \tr{\bar{B} S^\mu u_{\mu,i} B},
\end{equation}
which in contrast to the two-flavor case contains two isovector axial-vector coupling constants
$D$ and $F$, and in the present case of axionic HBCHPT the coupling constants $D^i$ to account for the
isoscalar interactions.

\subsection{General form of the axion-baryon coupling}\label{ch:generalshape}

For any process $B_B \to B_A + a$, where $B_A$ and $B_B$ denote arbitrary 
baryons and $a$ an
axion, the general Feynman rule for the vertex in the baryon rest frame has the form
\begin{equation}\label{eq:generalG1}
\raisebox{-0.8cm}{\includegraphics[height=1.8cm]{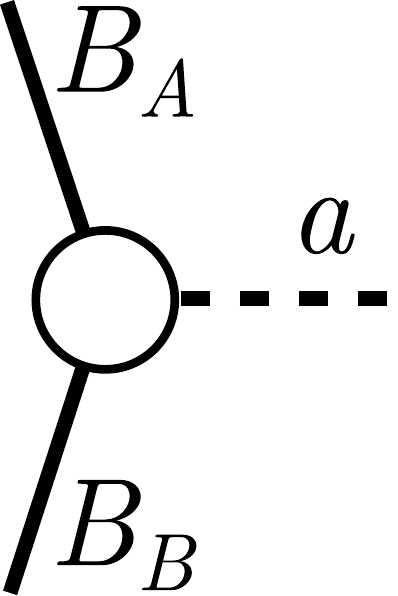}}\ = G_{aAB}\, \sdot{S}{q},
\end{equation}
where $q_\mu$ is the four-momentum of the outgoing axion. The coupling constant $G_{aAB}$ consists
of a power series in $1/f_a$
\begin{equation}\label{eq:generalG2}
G_{aAB} = -\frac{1}{f_a} g_{aAB} + \mathcal{O}\left(\frac{1}{f_a^2}\right).
\end{equation}
Due to the expected huge value of $f_a$, it is sufficient to determine the leading order term
$\propto  g_{aAB}/f_a$, where 
\begin{equation}\label{eq:generalcoupling}
 g_{aAB} = g^{(1)}_{aAB} + g^{(2)}_{aAB} + g_{aAB}^{(3)} + \dots
\end{equation}
can be expanded according to the power counting rules of HBCHPT. Here in particular
\begin{align}\label{eq:generalcoupling2}
\begin{split}
g^{(1)}_{aAB} & = g^\text{LO, tree}_{aAB}~,\\
g^{(2)}_{aAB} & = g^{1/m_B}_{aAB}~,\\
g^{(3)}_{aAB} & = g^\text{N\textsuperscript{2}LO, tree}_{aAB}+g^{1/m^2_B}_{aAB}+g^\text{LO, loop}_{aAB}~,
\end{split}
\end{align}
where $g^\text{LO}_{aAB}$ and $g^\text{N\textsuperscript{2}LO}_{aAB}$ refer to the contributions from the
leading and next-to-next-to-leading order chiral Lagrangian, i.\,e. from $\mathcal{L}_{\Phi B}^{(1)}$
and $\mathcal{L}_{\Phi B}^{(3)}$ of Eq.~\eqref{eq:lagrpower}, respectively. Furthermore, $g^{1/m_B}_{aAB}$ and $g^{1/m^2_B}_{aAB}$
refer to the $\mathcal{O}(m_B^{-1})$ and $\mathcal{O}(m_B^{-2})$ contributions from the expansion in the average baryon mass in the chiral limit $m_B$, which
additionally appear in HBCHPT (see section~\ref{ch:mBexpansion}). In this 
paper, we will
determine each of these contributions to $g_{aAB}$ up to  $\mathcal{O}\left( p^3\right)$.

As we work in the physical basis, $A$ and $B$ in $G_{aAB}$ and $g_{aAB}$ can be understood as SU(3)
indices running from $1$ to $8$ that directly relate to the physical baryon fields. $G_{a44}$, for
instance, thus gives the axion-proton coupling $G_{app}$, whereas $G_{a66}$ gives the axion-neutron
coupling $G_{ann}$ (see Appendix~\ref{appendix1}). In this basis, the matrix of $G_{aAB}$ is mostly
diagonal as we exclude flavor-changing axion-quark couplings, i.\,e. $G_{aAB}=0$ for $A\neq B$,
with the only exception of $G_{a38}=G_{a83}$, i.\,e. $G_{a\Sigma^0\Lambda}$, which is related
to the $\Sigma^0$-$\Lambda$ mixing.

\section{Determining the axion-baryon coupling}\label{ch:determiningcoupl}

\subsection{Leading order: tree level contributions}

From the Lagrangian Eq.~\eqref{eq:LagrLO}, we can determine the leading 
order axion-baryon
coupling by inserting\footnote{Note that in our previous work~\cite{Vonk:2020zfh}, there is a
typo in the corresponding SU(2) equations, i.\,e. Eqs. (3.16) and (3.21): 
$\tilde{u}_{\mu,i}$ is
not given by $\tilde{u}_{\mu,i}= c_i \frac{\partial_\mu a}{f_a} \tau_3$. The correct expression is
$\tilde{u}_{\mu,i}= c_i \frac{\partial_\mu a}{f_a} \mathbbm{1}$.}
\begin{align}
\begin{split}
  \Gamma_\mu & = 0~,\\ 
  u_\mu & =\frac{\partial_\mu a}{f_a} \left(c^{(3)} \lambda_3+c^{(8)}\lambda_8\right) ,\\
  u_{\mu, i} & =  c_i \frac{\partial_\mu a}{f_a} \mathbbm{1}~,
\end{split}
\end{align}
which gives
\begin{equation}\label{eq:LagrLOax}
\mathcal{L}_{\Phi B}^{(1),\text{int}} = \frac{g_{AB}}{f_a} \bar{B}_A \sdot{S}{\partial a} B_B~,
\end{equation}
where $A$ and $B$ are SU(3) indices in the physical basis (see Appendix~\ref{appendix1}). Here we have defined
\begin{align}\label{eq:gAB}
\begin{split}
g_{AB} = \frac{1}{2}\biggl\{ & D\left(c^{(3)} \tr{\lt{A}^\dagger \acom{\lambda_3}{\lt{B}}}+c^{(8)}
\tr{\lt{A}^\dagger \acom{\lambda_8}{\lt{B}}}\right)\\
& + F\left(c^{(3)} \tr{\lt{A}^\dagger \com{\lambda_3}{\lt{B}}}+c^{(8)} \tr{\lt{A}^\dagger
\com{\lambda_8}{\lt{B}}}\right) + 2 c_i D^i \delta_{AB}\biggr\}~. 
\end{split}
\end{align}
The matrix elements of $g_{AB}$ are given in Table~\ref{tab:gAB} in Appendix~\ref{appendix2}. As we are
working in the physical basis including the $\Sigma^0$-$\Lambda$ mixing, $g_{AB}$ actually includes
effects of $\mathcal{O}(p^2)$. The pure leading order axion-baryon coupling is found by setting
$\epsilon=0$, i.e. by replacing $\lt{A/B}\to\lh{A/B}$, see Eq.~\eqref{eq:lambdahut} of
Appendix~\ref{appendix1}. In this case, $\lambda_{3}$ and $\lambda_{8}$ can be replaced by
$\lh{3}$ and $\lh{8}$ so that the coupling can be expressed by means of the structure constants
$\fh{ABC}$ and $\dh{ABC}$:
\begin{equation}\label{eq:gAB0}
g_{AB}^0 = \left. g_{AB}\right|_{\epsilon=0} =c^{(3)}\left(D \dh{AB3}-F \fh{AB3}\right)
+ c^{(8)}\left(D \dh{AB8}-F \fh{AB8}\right)+ c_i D^i \delta_{AB}~.
\end{equation}
It is hence clear that $g^{(1)}_{aAB}$ in Eq.~\eqref{eq:generalcoupling} is simply given by
\begin{equation}
g^{(1)}_{aAB}= g_{AB}^0.
\end{equation}

\subsection{Expansion in the baryon mass}\label{ch:mBexpansion}

The Lagrangian containing the corrections due to the finite baryon masses 
$m_B$ in the heavy-baryon expansion up to $\mathcal{O}\left(p^3\right)$ corresponding to terms proportional
to $1/m_B$ and $1/m_B^2$ is given by~\cite{Muller:1996vy} 
\begin{equation}
\mathcal{L}^{1/m_B} = \bar{B}_A\left\{\frac{1}{2m_B}\gamma_0 \left[\mathcal{B}_{(1)}^{AC}\right]^\dagger \gamma_0
\mathcal{B}_{(1)}^{CB}-\frac{1}{4m_B^2}\gamma_0\left[\mathcal{B}_{(1)}^{AC}\right]^\dagger \gamma_0
\mathcal{A}_{(1)}^{CD} \mathcal{B}_{(1)}^{DB}\right\} B_B~.
\end{equation}
Here 
\begin{align}
\mathcal{A}^{AB}_{(1)} = \frac{1}{2}\biggl(\tr{\lt{A}^\dagger \com{i \sdot{v}{\mathcal{D}}}{\lt{B}}} &
+ D \tr{\lt{A}^\dagger \acom{\sdot{S}{u}}{\lt{B}}}\nonumber \\ &  + F \tr{\lt{A}^\dagger
\com{\sdot{S}{u}}{\lt{B}}} +D^i \tr{\lt{A}^\dagger \sdot{S}{u_i} \lt{B}}\biggr),
\end{align}
and
\begin{align}
\mathcal{B}^{AB}_{(1)} = \frac{1}{2}\biggl(\tr{\lt{A}^\dagger \com{i \gamma^\mu\mathcal{D}_\mu^\perp}{\lt{B}}}
& - \frac{D}{2} \tr{\lt{A}^\dagger \acom{\sdot{v}{u} \gamma_5}{\lt{B}}} \nonumber \\
& - \frac{F}{2}\tr{\lt{A}^\dagger \com{\sdot{v}{u} \gamma_5}{\lt{B}}} -\frac{D^i}{2}
\tr{\lt{A}^\dagger \sdot{v}{u_i} \gamma_5 \lt{B}}\biggr),
\end{align}
where for any four-vector $x_\mu$
\begin{equation}
x^\perp_\mu = v_\mu \sdot{v}{x} - x_\mu~.
\end{equation}
In the present case of axionic HBCHPT, we find
\begin{align}
\mathcal{A}^{AB}_{(1)} & = i \sdot{v}{\partial} \delta_{AB} + \frac{1}{f_a} g_{AB} \sdot{S}{\partial a}, \\
\mathcal{B}^{AB}_{(1)} & = i \gamma^\mu \partial_\mu^\perp \delta_{AB} - \frac{1}{2 f_a} g_{AB}
\sdot{v}{\partial a} \gamma_5~,
\end{align}
with $g_{AB}$ as defined in Eq.~\eqref{eq:gAB}, so that
\begin{align}
\mathcal{L}^{1/m_B} = \frac{1}{f_a}\bar{B}_A\biggl\{\frac{i g_{AB}}{2m_B} \acom{\sdot{S}{\partial}}{\sdot{v}{\partial a}} +\frac{g_{AB}}{4m_B^2}\bigl[ & -\partial^\mu\sdot{S}{\partial a}\partial_\mu + \sdot{v}{\partial}\sdot{S}{\partial a}\sdot{v}{\partial} \nonumber \\ & -\left(\acom{\sdot{S}{\partial}}{\sdot{v}{\partial a}}\sdot{v}{\partial} + \mathrm{h.c.}\right) \nonumber \\ & +\left(\sdot{S}{\partial}\sdot{\partial a}{\partial} 
+ \mathrm{h.c.}\right) \bigr]\biggr\} B_B~.
\end{align}
Let $p$ be the momentum of the incoming baryon, $p^\prime$ the momentum of the outgoing baryon,
and $\omega^{(\prime)}=\sdot{v}{p^{(\prime)}}$, then the resulting vertex Feynman rule reads
\begin{align}
\raisebox{-0.8cm}{\includegraphics[height=1.8cm]{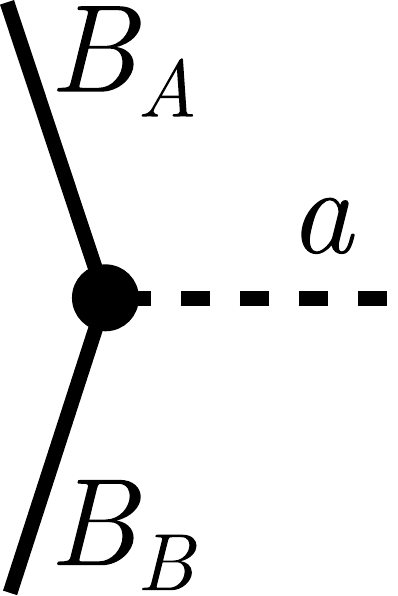}}\ = & -\frac{g_{AB}}{f_a}\left\{\frac{1}{2m_B} \left(\omega-\omega^\prime\right)-\frac{1}{4m_B^2} \left(\omega^2-{\omega^\prime}^2+\omega\omega^\prime-p^2\right) \right\} \sdot{S}{q}\nonumber\\ & +\frac{g_{AB}}{f_a}\left\{\frac{1}{m_B} \left(\omega-\omega^\prime\right)-\frac{1}{2m_B^2} \left(\omega^2-{\omega^\prime}^2-\frac{1}{2}\left(p^2-{p^\prime}^2\right)\right) \right\} \sdot{S}{p} ,
\end{align}
which is in complete analogy to the SU(2) case~\cite{Vonk:2020zfh}.
In the baryon rest frame with $p=0$, $\omega=0$, $v=\left(1,0,0,0\right)^\text{T}$,
and $\omega^\prime=\sdot{v}{p^\prime}=-\sdot{v}{q}=-q_0\ll m_B$, where $q_0$ is the relativistic energy
of the outgoing axion, we finally find (see Eqs.~\eqref{eq:generalcoupling} and \eqref{eq:generalcoupling2})
\begin{equation}\label{eq:massexpandresult}
g^\text{LO, tree}_{aAB}+g^{1/m_B}_{aAB}+g^{1/m^2_B}_{aAB} = g_{AB}\left\{1+\frac{q_0}{2m_B} +\frac{q_0^2}{4m_B^2} \right\}.
\end{equation}

\subsection{Next-to-next-to-leading order: Contributions from \texorpdfstring{{\boldmath$\chi_-$}}{X-}}
\label{ch:NNLOchim}

At next-to-next-to-leading order, there are several contributions that have to be considered. We can
differentiate terms with finite low-energy constants (LECs) that are $\propto \chi_-$, terms with
LECs having finite and ultraviolet (UV) divergent pieces $\propto \chi_+$, and terms proportional some LECs that
have no finite pieces serving as counter-terms to cancel UV divergences from the loop contributions.
We start with the former.

Up to $\mathcal{O}\left(1/f_a\right)$, $\chi_-$, see Eqs.~\eqref{eq:Mass:a} and \eqref{eq:chipm},
can be written as
\begin{equation}\label{eq:chim}
\chi_- = \frac{4i M_{\pi^\pm}^2}{f_a} \frac{z}{(1+z)^2} \left(1+\frac{w}{1+z}\right)^{-1} a \mathbbm{1}~,
\end{equation} 
where $M_{\pi^\pm}^2$ is the leading order mass of the charged pions given in Eq.~\eqref{eq:mesonmasses}.
The contributing terms of $\mathcal{L}_{\Phi B}^{(3)}$ are (here and in the following section, we
enumerate the LECs according to the list of terms in Ref.~\cite{Frink:2006hx})
\begin{align}
\mathcal{L}^{\chi_-} = & -i d_2 \left(\tr{\sdot{\partial\bar{B}}{S} {\chi_-}{B}}+\tr{\bar{B} {\chi_-}{\sdot{S}{\partial B}}} \right)\nonumber\\ &  -i d_3 \left(\tr{\sdot{\partial\bar{B}}{S} \tr{\chi_-}{B}}+\tr{\bar{B} 
\tr{\chi_-}{\sdot{S}{\partial B}}} \right).
\end{align}
Inserting Eq.~\eqref{eq:chim} yields
\begin{equation}\label{eq:xhimcontr}
\raisebox{-0.8cm}{\includegraphics[height=1.8cm]{BBa_tree}}\
= \frac{4 M_{\pi^\pm}^2}{f_a} \frac{z}{(1+z)^2} \left(1+\frac{w}{1+z}\right)^{-1}  \left(d_2 + 3d_3\right)
\delta_{AB} \sdot{S}{q}.
\end{equation}
The reason for writing the coupling in this way is to match it
to the corresponding terms in the SU(2) case, which are given by~\cite{Vonk:2020zfh}
\begin{equation}\label{eq:xhimcontrSU2}
\raisebox{-0.8cm}{\includegraphics[height=1.8cm]{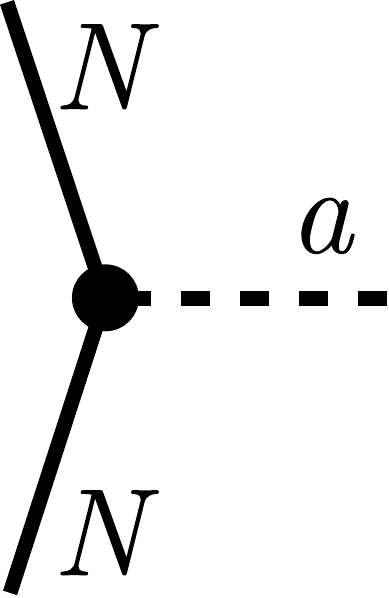}}\ = \frac{4 M_{\pi^\pm}^2}{f_a} \frac{z}{(1+z)^2}  \left( d_{18} + 2d_{19}\right) \sdot{S}{q},
\end{equation}
where $N$ is the nucleon field, and $d_{18}$ and $d_{19}$ are LECs from the next-to-next-to-leading order
SU(2) $\pi N$ Lagrangian~\cite{Fettes:1998ud}.
Apart from a substitution of the LECs, the main difference between the SU(3) case, Eq.~\eqref{eq:xhimcontr},
and the SU(2) case, Eq.~\eqref{eq:xhimcontrSU2}, is the explicit effect of the strange quark mass
$m_s$ in the factor $(1+w/(1+z))^{-1}$, which reduces to unity at $m_s\to\infty$, i.e. $w\to 0$.
In the SU(3) case, the finite value of $w$ accounts for a correction of about two percent. 

\subsection{Next-to-next-to-leading order: Contributions from \texorpdfstring{{\boldmath$\chi_+$}}{X+}}
\label{ch:NNLOchip}

From Eqs.~\eqref{eq:Mass:a} and \eqref{eq:chipm}, one can readily determine that
\begin{equation}
\chi_+ = 4B_0 \mathcal{M}_q + \mathcal{O}\left(f_a^{-2}\right).
\end{equation}
The relevant terms from $\mathcal{L}_{\Phi B}^{(3)}$ are~\cite{Frink:2006hx}
\begin{align}
\begin{split}
\mathcal{L}^{\chi_+} = & \phantom{+} d_{41}(\lambda) \left(\tr{\bar{B}S^\mu\com{u_\mu}{\com{\chi_+}{B}}}+\tr{\bar{B}S^\mu\com{\chi_+}{\com{u_\mu}{B}}}\right)\\
& + d_{42}(\lambda) \left(\tr{\bar{B}S^\mu\com{u_\mu}{\acom{\chi_+}{B}}}+\tr{\bar{B}S^\mu\acom{\chi_+}{\com{u_\mu}{B}}}\right)\\
& + d_{43}(\lambda) \left(\tr{\bar{B}S^\mu\acom{u_\mu}{\com{\chi_+}{B}}}+\tr{\bar{B}S^\mu\com{\chi_+}{\acom{u_\mu}{B}}}\right)\\
& + d_{44}(\lambda) \left(\tr{\bar{B}S^\mu\acom{u_\mu}{\acom{\chi_+}{B}}}+\tr{\bar{B}S^\mu\acom{\chi_+}{\acom{u_\mu}{B}}}\right)\\
& + d_{45}(\lambda) \tr{\bar{B}S^\mu\com{u_\mu}{B}}\tr{\chi_+} + d_{46}(\lambda) \tr{\bar{B}S^\mu\acom{u_\mu}{B}}\tr{\chi_+} \\
& + d_{47}(\lambda) \tr{\bar{B}S^\mu B}\tr{u_\mu \chi_+} + d_{43}^i(\lambda) \tr{\bar{B} \sdot{S}{u_i} \com{\chi_+}{B}} \\ 
& + d_{44}^i(\lambda) \tr{\bar{B} \sdot{S}{u_i} \acom{\chi_+}{B}}  + d^i_{46}(\lambda) \tr{\bar{B}\sdot{S}{u_i} B}\tr{\chi_+}.
\end{split}
\end{align}
The LECs depend on the scale $\lambda$ and are given by
\begin{equation}
d_{k}^{(i)} (\lambda) = d_{k}^{(i),r}(\lambda) + \frac{\beta^{(i)}_{k}}{F_p^2}L(\lambda)~,\qquad k=\{41,\dots,47\}~.
\end{equation}
In this equation, $d_{k}^{(i),r}(\lambda)$ refer to the renormalized LECs, and $L(\lambda)$ contains
the pole for spacetime dimension $d=4$,
\begin{equation}\label{eq:Ldivergence}
  L(\lambda)=\frac{\lambda^{d-4}}{(4\pi)^2}\left( \frac{1}{d-4}-\frac{1}{2}\left[\ln (4\pi)
    -\gamma+1\right]\right),
\end{equation}
where $\gamma$ is the Euler--Mascheroni constant. The $\beta$-functions are set to cancel the
divergences of the one-loop functional, as discussed below. We write the resulting tree-level vertex as
\begin{equation}\label{eq:xhipcontr}
\raisebox{-0.8cm}{\includegraphics[height=1.8cm]{BBa_tree}}\ = -\frac{4 M_{\pi^\pm}^2}{f_a}  \frac{z}{1+z} \hat{d}_{AB}(\lambda) \sdot{S}{q},
\end{equation}
where we have set
\begin{align}\label{eq:dABhut}
\begin{split}
\hat{d}_{AB}(\lambda) = \frac{1}{2 m_u} \Biggl( &  c^{(3)} \biggl\{ d_{41}(\lambda) \left(\tr{\lh{A}^\dagger \com{\lambda_{3}}{\com{\mathcal{M}_q}{\lh{B}}}}+\tr{\lh{A}^\dagger \com{\mathcal{M}_q}{\com{\lambda_{3}}{\lh{B}}}}\right)\\
&\qquad + d_{42}(\lambda) \left(\tr{\lh{A}^\dagger \com{\lambda_{3}}{\acom{\mathcal{M}_q}{\lh{B}}}}+\tr{\lh{A}^\dagger \acom{\mathcal{M}_q}{\com{\lambda_{3}}{\lh{B}}}}\right)\\
&\qquad + d_{43}(\lambda) \left(\tr{\lh{A}^\dagger \acom{\lambda_{3}}{\com{\mathcal{M}_q}{\lh{B}}}}+\tr{\lh{A}^\dagger \com{\mathcal{M}_q}{\acom{\lambda_{3}}{\lh{B}}}}\right)\\
&\qquad  + d_{44}(\lambda) \left(\tr{\lh{A}^\dagger \acom{\lambda_{3}}{\acom{\mathcal{M}_q}{\lh{B}}}}+\tr{\lh{A}^\dagger \acom{\mathcal{M}_q}{\acom{\lambda_{3}}{\lh{B}}}}\right)\\
&\qquad + d_{45}(\lambda) \tr{\lh{A}^\dagger\com{\lambda_{3}}{\lh{B}}}\tr{\mathcal{M}_q} + d_{46}(\lambda) \tr{\lh{A}^\dagger\acom{\lambda_{3}}{\lh{B}}}\tr{\mathcal{M}_q}\\
&\qquad  + 2 d_{47}(\lambda) \delta_{AB} \tr{\lambda_{3} \mathcal{M}_q}\biggr\}\\
& +c^{(8)}\biggl\{ d_{41}(\lambda) \left(\tr{\lh{A}^\dagger \com{\lambda_{8}}{\com{\mathcal{M}_q}{\lh{B}}}}+\tr{\lh{A}^\dagger \com{\mathcal{M}_q}{\com{\lambda_{8}}{\lh{B}}}}\right)\\
&\qquad + d_{42}(\lambda) \left(\tr{\lh{A}^\dagger \com{\lambda_{8}}{\acom{\mathcal{M}_q}{\lh{B}}}}+\tr{\lh{A}^\dagger \acom{\mathcal{M}_q}{\com{\lambda_{8}}{\lh{B}}}}\right)\\
& \qquad + d_{43}(\lambda) \left(\tr{\lh{A}^\dagger \acom{\lambda_{8}}{\com{\mathcal{M}_q}{\lh{B}}}}+\tr{\lh{A}^\dagger \com{\mathcal{M}_q}{\acom{\lambda_{8}}{\lh{B}}}}\right)\\
&\qquad + d_{44}(\lambda) \left(\tr{\lh{A}^\dagger \acom{\lambda_{8}}{\acom{\mathcal{M}_q}{\lh{B}}}}+\tr{\lh{A}^\dagger \acom{\mathcal{M}_q}{\acom{\lambda_{8}}{\lh{B}}}}\right)\\
&\qquad + d_{45}(\lambda) \tr{\lh{A}^\dagger\com{\lambda_{8}}{\lh{B}}}\tr{\mathcal{M}_q} + d_{46}(\lambda) \tr{\lh{A}^\dagger\acom{\lambda_{8}}{\lh{B}}}\tr{\mathcal{M}_q}\\
&\qquad + 2 d_{47}(\lambda) \delta_{AB} \tr{\lambda_{8} \mathcal{M}_q}\biggr\} \\
& +c_i \biggl\{d_{43}^i(\lambda) \tr{\lh{A} \com{\mathcal{M}_q}{\lh{B}}}+d_{44}^i(\lambda) \tr{\lh{A} \acom{\mathcal{M}_q}{\lh{B}}}\\
& \qquad + 2 d_{46}^i(\lambda) \delta_{AB} \tr{\mathcal{M}_q} \biggr\}\Biggr).
\end{split}
\end{align}

The matrix elements of $\hat{d}_{AB}$ are given in Eqs.~\eqref{eq:d11}--\eqref{eq:d83} of
Appendix~\ref{appendix2}.

\vfill

\subsection{Next-to-next-to-leading order: Counter-terms}
\label{ch:NNLOct}

The counter-terms needed for the renormalization have been worked out in Ref.~\cite{Muller:1996vy}.
For the present case, we need the terms $i=\{36,37,38,39\}$ from this paper, which are given by
\begin{align}
\begin{split}
\mathcal{L}^\text{c.t.} = & \phantom{+} d_{36}(\lambda) \tr{\sdot{v}{\partial\bar{B}}\acom{\sdot{S}{u}}{\sdot{v}{\partial B}}}\\
& + d_{37}(\lambda) \tr{\sdot{v}{\partial\bar{B}}\com{\sdot{S}{u}}{\sdot{v}{\partial B}}}\\
& + d_{38}(\lambda) \tr{\bar{B}\acom{\com{\sdot{v}{\partial}}{\com{\sdot{v}{\partial}}{\sdot{S}{u}}}}{B}}\\
&  + d_{39}(\lambda)\tr{\bar{B}\com{\com{\sdot{v}{\partial}}{\com{\sdot{v}{\partial}}{\sdot{S}{u}}}}{B}}\\
& + d^i_{36}(\lambda) \tr{\sdot{v}{\partial\bar{B}}\sdot{S}{u_i}\sdot{v}{\partial B}}\\
& + d_{38}^i(\lambda) \tr{\bar{B} \com{\sdot{v}{\partial}}{\com{\sdot{v}{\partial}}{\sdot{S}{u_i}}} B},
\end{split}
\end{align}
where we have added two terms in order to account for the isoscalar interactions.
The LECs have no finite part, i.e.
 \begin{equation}
   d_{k}^{(i)} (\lambda) =\frac{\beta^{(i)}_{k}}{F_p^2} L(\lambda)~,\qquad k=\{36,\dots,39\}~.
\end{equation}
Their contribution to the tree-level vertex is given by
\begin{equation}\label{eq:ctcontr}
\raisebox{-0.8cm}{\includegraphics[height=1.8cm]{BBa_tree}}\ = -\frac{1}{f_a} \hat{d}^{c.t.}_{AB} (\lambda) \sdot{S}{q},
\end{equation}
with
\begin{align}
\begin{split}
\hat{d}^{c.t.}_{AB}(\lambda) = & \phantom{+} \left( d_{36}(\lambda) \omega\omega^\prime
-d_{38}(\lambda)\left(\omega-\omega^\prime\right)^2\right)\left(c^{(3)} \dh{AB3}+c^{(8)}\dh{AB8}\right)\\
& - \left( d_{37}(\lambda) \omega\omega^\prime -d_{39}(\lambda)\left(\omega-\omega^\prime\right)^2\right)
\left(c^{(3)} \fh{AB3}+c^{(8)}\fh{AB8}\right)\\
& + \left( d_{36}^i(\lambda) \omega\omega^\prime -d_{38}^i(\lambda)\left(\omega-\omega^\prime\right)^2\right)
c_i  \delta_{AB}~.
\end{split}
\end{align}

\subsection{Loops}\label{ch:loops}
\begin{figure}[t]
\includegraphics[width=0.6\textwidth]{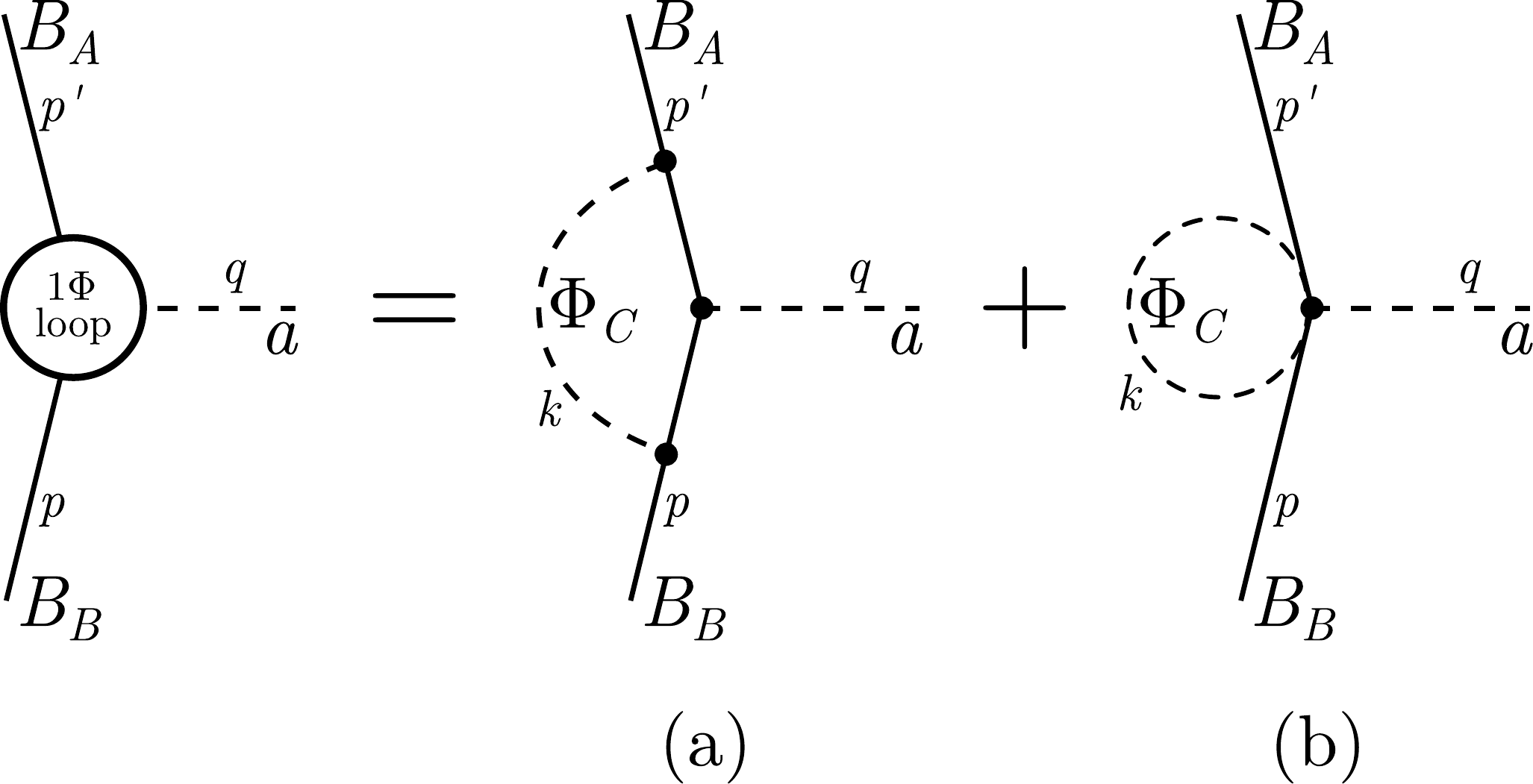}
\caption{Non-vanishing meson loop contributions to $B_B\to B_A+a$.}
\label{fig:oneloop}
\end{figure}
There are only two non-vanishing single meson loops that contribute to the $\mathcal{O}(p^3)$
axion-baryon vertex at $\mathcal{O}(1/f_a)$, which are shown in Fig.~\ref{fig:oneloop}.
In this figure, mesons are identified by means of the SU(3) index $C$ in the physical basis
including the $\pi^0$-$\eta$ mixing, cf. Eqs.~\eqref{eq:mesons} and \eqref{eq:mesonsexplicit}.
Note that the potential diagrams with a $a\Phi BB$ vertex and a meson line
connected to one baryon leg vanish because $\sdot{v}{S} = 0$, see Eq.~\eqref{eq:spinoperator}, as has
been shown in Ref.~\cite{Vonk:2020zfh} for the corresponding diagrams in the SU(2) case, which
have the same topology.

\subsubsection{Diagram (a)}

Using the leading order meson-baryon vertex rule, one finds
\begin{align}\label{eq:diagrama}
\begin{split}
  \text{(a)} & = \frac{1}{f_aF_p^2} \sum_C g_{ABC}^{(a)} S^\mu \sdot{S}{q} S^\nu \frac{1}{i}\int
  \frac{\mathrm{d}^dk}{(2\pi)^d}\, \frac{k_\mu k_\nu}{(k^2-M_{\Phi_C}^2+i\eta)(\omega^\prime
    - v\cdot k+i\eta)(\omega - v\cdot k+i\eta)}\\
    & = \frac{1}{6f_a}\frac{1}{(4\pi F_p)^2} \sum_{C} g_{ABC}^{(a)} \Biggl\{-M_{\Phi_C}^2
 \\ & \qquad +\frac{1}{\omega-\omega^\prime}\biggl(\omega^3-{\omega^\prime}^3  +2\left[\left(M_{\Phi_C}^2-\omega^2\right)^\frac{3}{2}\arccos\frac{-\omega}{M_{\Phi_C}} -\left(M_{\Phi_C}^2
    -{\omega^\prime}^2\right)^\frac{3}{2}\arccos\frac{-\omega^\prime}{M_{\Phi_C}}\right]\biggr)\\ & \qquad+\left(3M_{\Phi_C}^2-2\left(\omega-\omega^\prime\right)^2
  -6\omega\omega^\prime \right)\left((4\pi)^2 L(\lambda)+\ln \frac{M_{\Phi_C}}{\lambda}\right)\Biggr\} \sdot{S}{q},
\end{split}
\end{align}
where we have applied dimensional regularization and used the properties in Eq.~\eqref{eq:spinoperator}.
Moreover, we have defined
\begin{align}
\begin{split}
g_{ABC}^{(a)} = \frac{1}{4} \sum_{D,E} & \left(D \tr{\lh{A}^\dagger\acom{\lt{C}}{\lh{E}}}+F \tr{\lh{A}^\dagger\com{\lt{C}}{\lh{E}}}\right) g_{ED}^0 \\ & \qquad \times  \left(D \tr{\lh{D}^\dagger\acom{\lt{C}^\dagger}{\lh{B}}}+F \tr{\lh{D}^\dagger\com{\lt{C}^\dagger}{\lh{B}}}\right),
\end{split}
\end{align}
in order to handle all meson loops for all baryons $B_A$ and $B_B$ at the 
same time. Here $g_{ED}^0$ refers to Eq.~\eqref{eq:gAB0}, the leading order axion-baryon coupling.
Equation~\eqref{eq:diagrama} contains UV divergences, which will be treated below in section~\ref{ch:renorm}.
The expression for diagram~(a) can be simplified by considering the baryon rest frame
and expanding around $q_0 \ll M_{\Phi_C}$ for all mesons $\Phi_C$, which yields
\begin{align}\label{eq:diagramaapprox}
\begin{split}
  \text{(a)} & = \frac{1}{6f_a} \sum_{C}  g_{ABC}^{(a)} \left(\frac{M_{\Phi_C}}{4\pi F_p}\right)^2 \Biggl\{1+\frac{3\pi}{2}\frac{q_0}{M_{\Phi_C}}-\frac{5}{3}\left(\frac{q_0}{M_{\Phi_C}}\right)^2
\\ & \qquad+\left(3-2\left(\frac{q_0}{M_{\Phi_C}}\right)^2\right)\left((4\pi)^2 L(\lambda)+\ln \frac{M_{\Phi_C}}{\lambda}\right)\Biggr\} \sdot{S}{q} .
\end{split}
\end{align}

\subsubsection{Diagram (b)}

Expanding $u$ up to order $\Phi^2$ yields
\begin{equation}
u_\mu = -\frac{\partial_\mu a }{8f_aF_p^2} \left(c^{(3)} \com{\Phi}{\com{\Phi}{\lambda_3}}+c^{(8)}
\com{\Phi}{\com{\Phi}{\lambda_8}}\right),
\end{equation}
so the vertex rule for the $aB_AB_B\Phi_C\Phi_C$-vertex in the physical basis derived from
$\mathcal{L}_{\Phi B}^{(1)}$ can be written as
\begin{equation}
\frac{1}{f_a F_p^2} g_{ABC}^{(b)} \sdot{S}{q},
\end{equation}
where we have defined the coupling constant
\begin{align}\label{eq:gaMMBB}
\begin{split}
g_{ABC}^{(b)} = \frac{1}{8} \biggr\{ & c^{(3)}\left(D \tr{\lh{A}^\dagger\acom{\com{\lt{C}}{\com{\lt{C}^\dagger}
{\lambda_3}}}{\lh{B}}}+F \tr{\lh{A}^\dagger\com{\com{\lt{C}}{\com{\lt{C}^\dagger}{\lambda_3}}}{\lh{B}}}\right)\\
& + c^{(8)}\left(D \tr{\lh{A}^\dagger\acom{\com{\lt{C}}{\com{\lt{C}^\dagger}{\lambda_8}}}{\lh{B}}}
+F \tr{\lh{A}^\dagger\com{\com{\lt{C}}{\com{\lt{C}^\dagger}{\lambda_8}}}{\lh{B}}}\right)\biggl\}\, .
\end{split}
\end{align}
In fact, this vertex and thus diagram~(b) is independent of the mixing angle $\epsilon$ so one might
as well substitute $\lt{C}\to \lh{C}$ in Eq.~\eqref{eq:gaMMBB}. Therefore, one can also express the
coupling constant by means of the structure constants defined in Eqs.~\eqref{eq:commutatorrelations}
and \eqref{eq:lambdahut},
\begin{equation}\label{eq:gaMMBB2}
g_{ABC}^{(b)} = \frac{1}{4} \sum_{D,E} \fh{ECD} \left( c^{(3)} \fh{3CD} 
\left[D \dh{AEB} + F \fh{AEB} \right]
+c^{(8)} \fh{8CD} \left[D \dh{AEB} + F \fh{AEB} \right] \right).
\end{equation}
The loops of diagram~(b) for all mesons $\Phi_C$ can then be calculated as
\begin{align}
\begin{split}
\text{(b)} & = \frac{1}{2 f_a F_p^2}\sdot{S}{q} \sum_C g_{ABC}^{(b)} \frac{1}{i}
\int\frac{\mathrm{d}^dk}{(2\pi)^d}\,
\frac{1}{k^2-M_{\Phi_C}^2+i\eta} \\ &  = -\frac{1}{f_a F_p^2}\sdot{S}{q} \sum_C g_{ABC}^{(b)} M^2_{\Phi_C}
\left(L(\lambda)+\frac{1}{(4\pi)^2} \ln\frac{M_{\Phi_c}}{\lambda}\right). 
\end{split}
\end{align}

\subsubsection{Renormalization}\label{ch:renorm}

The divergences appearing in the meson loop calculations in dimensional regularization are
canceled by setting appropriate $\beta$ functions for the LECs appearing in $\hat{d}_{AB}(\lambda)$,
Eq.~\eqref{eq:dABhut}, and $\hat{d}^{c.t.}_{AB}(\lambda)$, Eq.~\eqref{eq:ctcontr},
\begin{align}
\beta_{36} & = 2 D \left(D^2+3F^2 \right) , & \beta^i_{36} & = -\frac{4}{3} D^i \left(13D^2+9F^2 \right) ,
\nonumber  \\
\beta_{37} & = \frac{2}{3} F \left(5D^2-9F^2 \right) , & \beta_{38} & = 
-\frac{2}{3} D \left(D^2+3F^2 \right) ,
\nonumber \\
\beta^i_{38} & = \frac{4}{9} D^i \left(13D^2+9F^2 \right) , & \beta_{39} & = -\frac{2}{9} F \left(5D^2-9F^2
\right) ,\nonumber \\
\beta_{41} & = -\frac{1}{48} D \left(9D^2+7F^2-9 \right) , & \beta_{42} 
& = -\frac{3}{16} F \left(D^2-F^2-1
\right) ,\nonumber \\
\beta_{43} & = -\frac{1}{48} F \left(7D^2+9F^2-9 \right) , & \beta^i_{43} & = -\frac{5}{6} DFD^i~,\nonumber \\
\beta_{44} & = -\frac{1}{48} D \left(23D^2+9F^2-9 \right) , & \beta^i_{44} & = -\frac{1}{4} D^i
\left(D^2-3F^2\right) ,\nonumber \\
\beta_{45} & = -\frac{1}{36} F \left(D^2-9F^2-9 \right) , & \beta_{46} & = \frac{1}{36} D \left(17D^2-9F^2+9
\right) , \nonumber \\
\beta^i_{46} & = \frac{1}{18} D^i \left(13D^2+9F^2\right) & \beta_{47} & = \frac{1}{2} D
\left(D^2+3F^2-1 \right) ,
\end{align}
of which the $\beta_k$'s are in accordance with the ones given in Ref.~\cite{Muller:1996vy}, and the $\beta^i_k$'s have been worked out here for the first time. With that, the full renormalized $\mathcal{O}(p^3)$ contribution reads
\begin{equation}\label{eq:NNLOloop}
g^\text{N\textsuperscript{2}LO, tree}_{aAB} + g_{aAB}^\text{LO, loop} = 
 \frac{ 4 M_{\pi^\pm}^2 z}{1+z}\left(\hat{d}^r_{AB} (\lambda) -\frac{d_2 + 3d_3}{1+z+w} \delta_{AB}\right) -\frac{1}{27} g_{AB}^{\text{loop},r}  +g_{AB}^{\text{loop,sc}} (\lambda) ~,
\end{equation}
where we have neglected terms of $\mathcal{O}\left(q_0/M_{\Phi_C}\right)$. Moreover, $\hat{d}^r_{AB}$ refers
to Eq.~\eqref{eq:dABhut} with renormalized LECs, and
\begin{align}
g_{AB}^{\text{loop},r} & = \frac{9}{2} \sum_C g_{ABC}^{(a)} \left(\frac{M_{\Phi_C}}{4\pi F_p}\right)^2 ~, \label{eq:gABloopr} \\
g_{AB}^{\text{loop,sc}} (\lambda) & = \sum_C \left(-\frac{1}{2} g_{ABC}^{(a)} +g_{ABC}^{(b)}\right) \left(\frac{M_{\Phi_C}}{4\pi F_p}\right)^2
\ln\frac{M_{\Phi_C}}{\lambda}~.
\end{align}
The matrix elements of $g_{AB}^{\text{loop},r}$ are given in Eqs.~\eqref{eq:gloop11}--\eqref{eq:gloop38}
of Appendix~\ref{appendix2}.

\section{Results}\label{ch:results}

\subsection{Leading order axion-baryon coupling}\label{ch:baryoncoupl}

Using the nucleon matrix elements $\Delta q$ defined by  $s^\mu\Delta q=\langle p | \bar{q}\gamma^\mu\gamma_5
q|p\rangle$, $s^\mu$ being the spin of the proton, we set
\begin{align}\label{eq:DFchoice}
(D+F) & = g_A = \Delta u-\Delta d~, &
-(D-3F)  &= \Delta u+\Delta d-2\Delta s~, \\
D^1 & = \Delta u+\Delta d+\Delta s~, &
D^i &= \Delta q_i~,\text{ for } i=\{2,3,4\}, ~ \Delta q_i=\{\Delta c,\Delta b,\Delta t\}~, \nonumber
\end{align}
where the respective baryon matrix elements are related to the ones from the nucleons by flavor symmetry. 
From now on, we neglect terms $\propto \{\Delta c,\Delta b,\Delta t\}$, 
which basically represent
sea quark effects beyond the numerical uncertainties of the dominant contributions of the up, down,
and strange quarks (at least in the standard DFSZ scenario, where the couplings to heavy quarks are
of the same order as the couplings to the light quarks; if alternatively the couplings $X_c$, $X_b$,
and $X_t$ or only one or two of them were much stronger than $X_u$, $X_d$, and $X_s$, these sea
quark terms could not be ignored any longer). Inserting this into the result for the leading
order axion-baryon coupling constant $g^{(1)}_{aAB}$, Eq.~\eqref{eq:gAB0} 
(see also Table~\ref{tab:gAB} in
Appendix~\ref{appendix2}), we find  
\begin{align}
\begin{split}
g^{(1)}_{a\Sigma^+\Sigma^+} & = -\frac{\Delta u+z\Delta s+w\Delta d}{1+z+w}+\Delta u X_u + \Delta s X_d+ \Delta d X_s ~,\\
g^{(1)}_{a\Sigma^-\Sigma^-} & = -\frac{\Delta s+z\Delta u+w\Delta d}{1+z+w}+\Delta s X_u + \Delta u X_d+ \Delta d X_s ~,\\
g^{(1)}_{a\Sigma^0\Sigma^0} & = -\frac{\frac{\Delta u+\Delta s}{2}(1+z)+w\Delta d}{1+z+w}+\frac{\Delta u+\Delta s}{2} \left( X_u+X_d\right) + \Delta d X_s ~,\\
g^{(1)}_{app} & = -\frac{\Delta u+z\Delta d+w\Delta s}{1+z+w}+\Delta u X_u + \Delta d X_d+ \Delta s X_s ~,\\
g^{(1)}_{a\Xi^-\Xi^-} & = -\frac{\Delta s+z\Delta d+w\Delta u}{1+z+w}+\Delta s X_u + \Delta d X_d+ \Delta u X_s ~,\\
g^{(1)}_{ann} & = -\frac{\Delta d+z\Delta u+w\Delta s}{1+z+w}+\Delta d X_u + \Delta u X_d+ \Delta s X_s ~,\\
g^{(1)}_{a\Xi^0\Xi^0} & = -\frac{\Delta d+z\Delta s+w\Delta u}{1+z+w}+\Delta d X_u + \Delta s X_d+ \Delta u X_s ~,\\
g^{(1)}_{a\Lambda\Lambda} & =-\frac{\frac{\Delta u+4\Delta d+\Delta s}{6} \left(1+z \right)+\frac{2\Delta u-\Delta d+2\Delta s}{3} w}{1+z+w}\\ & 
\qquad \qquad +\frac{\Delta u+4\Delta d+\Delta s}{6} \left(X_u+X_d \right)+\frac{2\Delta u-\Delta d+2\Delta s}{3} X_s ~,\\
g^{(1)}_{a\Sigma^0\Lambda} & = -\frac{\frac{\Delta u-2\Delta d+\Delta s}{2\sqrt{3}} \left(1-z \right)}{1+z+w} +\frac{\Delta u-2\Delta d+\Delta s}{2\sqrt{3}} \left(X_u-X_d\right) .
\end{split}
\end{align}
In particular, $g^{(1)}_{app}$ and $g^{(1)}_{ann}$ are exactly the same as in the SU(2) case.
Using~\cite{Aoki:2019cca}
\begin{equation}
\Delta u  = 0.847(50)~,\qquad \Delta d = -0.407(34)~, \qquad \Delta s 
 = -0.035(13)~,
\end{equation}
which correspond to
\begin{align}
\begin{split}
D & = \frac{1}{2}\Delta u - \Delta d +\frac{1}{2}\Delta s = 0.813(43)~,\\
F & = \frac{1}{2}\Delta u -\frac{1}{2}\Delta s = 0.441(26)~,\\
D^1& = 0.405(62)~,\\
\end{split}
\end{align}
and~\cite{Aoki:2019cca}
\begin{equation}
z=0.485(19)~,\qquad w=0.025(1)~,
\end{equation}
we obtain
\begin{align}\label{eq:gLO}
\begin{split}
g^{(1)}_{a\Sigma^+\Sigma^+} & = -0.543(34)+0.847(50) X_u -0.035(13) X_d 
-0.407(34) X_s ~,\\
g^{(1)}_{a\Sigma^-\Sigma^-} & = -0.242(21)-0.035(13) X_u +0.847(50) X_d 
-0.407(34) X_s ~,\\
g^{(1)}_{a\Sigma^0\Sigma^0} & = -0.396(25)+0.417(25) X_u + 0.395(25)X_d 
-0.407(35) X_s ~,\\
g^{(1)}_{app} & = -0.430(36) + 0.847(50)X_u -0.407(34) X_d -0.035(13) X_s ~,\\
g^{(1)}_{a\Xi^-\Xi^-} & =  \phantom{-}0.140(15)-0.035(13) X_u -0.407(34) X_d +0.847(50) X_s  ~,\\
g^{(1)}_{ann} & =  -0.002(30)-0.407(34) X_u + 0.847(50)X_d -0.035(13) X_s ~,\\
g^{(1)}_{a\Xi^0\Xi^0} & =  \phantom{-}0.267(23)-0.407(34) X_u -0.035(13) X_d +0.847(50) X_s  ~,\\
g^{(1)}_{a\Lambda\Lambda} & =\phantom{-}0.126(25) -0.147(25) X_u -0.125(25) X_d +0.677(35) X_s  ~,\\
g^{(1)}_{a\Sigma^0\Lambda} & = -0.153(10) +0.463(25) X_u -0.476(25) X_d 
+ 0.013(1) X_s ~,
\end{split}
\end{align}
where we also considered corrections from non-vanishing mixing angle $\epsilon$ related to isospin
breaking in the cases of the $\Sigma^0$ and the $\Lambda$ (which is why there also appears a term
$\propto X_s$ in $g^{(1)}_{a\Sigma^0\Lambda}$). In Table~\ref{tab:gABaxbar}, we list the results for the
KSVZ axion, where $X_q=0$, and the DFSZ axion, where the axion-quark couplings $X_q$ depend on the
angle $\beta$ related to the VEVs of the involved Higgs doublets (see above,
Eq.~\eqref{eq:couplignconstantsmodeldepending}). In the KSVZ model, the strongest couplings are hence
to be expected for the $\Sigma^+$ and the proton, which is also true for the DFSZ model at small
values of $\sin^2\beta$ (in this region, also the $\Xi^0$ shows a considerably large coupling with an
opposite sign). As noted already in many previous works, the axion-neutron coupling in some scenarios is strongly
suppressed and might even vanish in the KSVZ model and the DFSZ model at $\sin^2\beta\approx 2/3$
(corresponding to $x=1/\sqrt{2}$, where $x$ is the ratio of the VEVs of 
the two Higgs doublets).
The axion-neutron coupling is also the only baryon conserving coupling that might vanish in the DFSZ model,
as it is the only one that changes its sign when varying $\sin^2\beta$ from zero to unity. At $\sin^2\beta=1$,
also the $\Sigma^0$-$\Lambda$ mixing vertex disappears. At the same value 
of $\sin^2\beta$, the
couplings are somehow ``harmonized", i.e. the couplings of the axion to particles of the same
strangeness $S$ are approximately the same,  which is due to flavor symmetry. In case of the neutron
and the proton with $S=0$ one then has $g_{aAB}^{(1),\sin^2\beta=1}\approx -0.14$, in case of the
$\Sigma$ particles with $S=1$, one has $g_{aAB}^{(1),\sin^2\beta=1}\approx -0.26$, and in case of the
two $\Xi$ baryons with $S=2$, one has $g_{aAB}^{(1),\sin^2\beta=1}\approx 0.13$.  The difference among the
particles with the same $S$ can be determined as being always
\begin{equation}\label{eq:deltag}
\Delta g_{aAB}^{(1),\sin^2\beta=1} = \frac{2-4z-w}{3(1+z+w)} \left(\Delta q_1 - \Delta q_2\right) \approx 0.008
\left(\Delta q_1 - \Delta q_2\right) ,
\end{equation}
where $AB$ here denotes particles of the same strangeness, and $\Delta q_1$ and $\Delta q_2$ depends
on the quark content of these particles, i.e.  $\left(\Delta q_1 - \Delta 
q_2\right)=\left(\Delta s
- \Delta u\right)$ in the case of the $\Sigma$ baryons,  $\left(\Delta q_1 - \Delta q_2\right)=
\left(\Delta d- \Delta u\right)$ in the case of the nucleons, and  $\left(\Delta q_1 - \Delta q_2\right)
=\left(\Delta d- \Delta s\right)$ in the case of the $\Xi$ particles (note that in cases with $\Sigma^0$
an additional factor $1/2$ and corrections from $\epsilon\neq 0$ appear in Eq.~\eqref{eq:deltag}).
\begin{table}
\begin{tabular}{c|c|c|c|c|c}
\hline
\multirow{3}{*}{Process} & \multicolumn{5}{|c}{$g^{(1)}_{aAB}$} \\ \cline{2-6}
& \multirow{2}{*}{KSVZ} & \multicolumn{4}{|c}{DFSZ} \\ \cline{3-6}
& & general & ${\sin^2\beta=0}$ & ${\sin^2\beta=\tfrac{2}{3}}$ & ${\sin^2\beta=1}$\\
\hline
$\Sigma^+ \to\Sigma^+ + a$ &  $-0.543(34)$ &  $- 0.690(36) + 0.430(21) \sin^2\beta$ &$- 0.690(36)$ &$- 0.404(36)$ & $- 0.261(38)$\\
$\Sigma^- \to\Sigma^- + a$ & $-0.242(21)$ & $- 0.095(29) - 0.158(21) \sin^2\beta$ &$- 0.095(29)$ & $- 0.201(23)$ & $-0.254(22)$\\
$\Sigma^0 \to\Sigma^0 + a$ & $-0.396(25)$ & $- 0.400(29) + 0.143(12) \sin^2\beta$ & $- 0.400(29)$ & $- 0.305(27)$ & $- 0.257(27)$ \\
$p \to p + a$ & $-0.430(36)$ & $- 0.577(38) + 0.430(21) \sin^2\beta$ &$- 0.577(38)$ & $- 0.291(38)$ & $- 0.147(39)$\\
$\Xi^- \to\Xi^- + a$ &  $\phantom{-}0.140(15)$ & $\phantom{-} 0.287(25) - 
0.158(21) \sin^2\beta$ &$\phantom{-} 0.287(25)$ & $\phantom{-} 0.181(17)$ 
& $\phantom{-} 0.128(16)$ \\
$n \to n + a$ & $-0.002(30)$ & $\phantom{-} 0.269(34) - 0.406(21) \sin^2\beta$ & $\phantom{-} 0.269(34)$ & $- 0.002(31)$ & $- 0.138(22)$ \\
$\Xi^0 \to\Xi^0 + a$ & $\phantom{-}0.267(23)$ & $\phantom{-} 0.531(29) - 0.406(21) \sin^2\beta$ & $\phantom{-} 0.531(29)$ & $\phantom{-} 0.267(25)$ & $\phantom{-} 0.131(26)$ \\
$\Lambda \to\Lambda + a$ & $\phantom{-}0.126(25)$ & $\phantom{-} 0.310(29) - 0.233(12) \sin^2\beta$ & $\phantom{-} 0.310(29)$ & $\phantom{-} 0.155(27)$& $\phantom{-} 0.077(26)$ \\ \hline
$\Sigma^0 \to\Lambda + a$ & \multirow{2}{*}{$-0.153(10) $} & \multirow{2}{*}{$- 0.308(13) + 0.309(16) \sin^2\beta$} & \multirow{2}{*}{$- 0.308(13)$} & \multirow{2}{*}{$- 0.102(11)$} & \multirow{2}{*}{$\phantom{-} 0.000(13)$} \\
$\Lambda \to\Sigma^0 + a$ &  &  & & & \\ \hline
\end{tabular}
\caption{Leading order axion-baryon couplings $g^{(1)}_{aAB}$ for the KSVZ axion and the DFSZ axion.}
\label{tab:gABaxbar}
\end{table}

\subsection{Loop corrections and estimation of the NNLO LECs}\label{ch:NNLO}

As stated already, the results for the leading order axion-nucleon coupling in the $N_f=3$ case are entirely in line with the results of the $N_f=2$ case, which is also true for the $\mathcal{O}\left(p^2\right)$ and 
$\mathcal{O}\left(p^3\right)$ corrections stemming from the expansion in $1/m_B$ that appear in the non-relativistic heavy baryon limit, with the only exception that the nucleon mass in the chiral limit $m_0$ appearing in the SU(2) case is substituted by the average baryon mass in the chiral 
limit $m_B$. In the limit of soft axions, i.e. $q_0\to 0$, these terms, see Eq.~\eqref{eq:massexpandresult}, rapidly vanish.

The more significant corrections stem from the one-meson loop contributions, Eq~\eqref{eq:NNLOloop}. For the calculation of the corresponding matrix elements, we use the physical meson masses and decay constants~\cite{Zyla:2020zbs,Kolesar:2019sux},
$M_{\pi^\pm}  = 139.57\,\text{MeV},  M_{\pi^0}  = 134.98\,\text{MeV},
M_{K^\pm}  = 493.68\,\text{MeV},  M_{K^0}  = 497.61\,\text{MeV},
M_{\eta}  = 547.86\,\text{MeV},  F_{\pi}  = 92.1(6)\,\text{MeV},
F_{K}  = 110.3(5)\,\text{MeV},$ and  $F_{\eta}  = 118(9)\,\text{MeV}.$
Inserting this numerical input, yields
\begin{align}\label{eq:gloopcontr}
\begin{split}
g^\text{loop}_{a\Sigma^+\Sigma^+} & = \phantom{-}0.096(7)-0.148(10) X_u -0.003(5) X_d +0.202(16) X_s ~,\\
g^\text{loop}_{a\Sigma^-\Sigma^-} & = \phantom{-}0.046(6)-0.003(5) X_u -0.148(10) X_d +0.202(16) X_s ~,\\
g^\text{loop}_{a\Sigma^0\Sigma^0} & = \phantom{-}0.072(7)-0.076(8) X_u - 0.076(8)X_d +0.202(16) X_s ~,\\
g^\text{loop}_{app} & = 				\phantom{-}0.046(5) - 0.119(11)X_u +0.098(7) X_d +0.064(8) X_s ~,\\
g^\text{loop}_{a\Xi^-\Xi^-} & =  			-0.112(8)+0.086(8) X_u +0.182(16) X_d -0.211(16) X_s  ~,\\
g^\text{loop}_{ann} & =  					-0.028(5)+0.098(7) X_u - 0.119(11)X_d +0.064(8) X_s ~,\\
g^\text{loop}_{a\Xi^0\Xi^0} & =  			-0.145(11)+0.182(16) X_u +0.086(8) X_d -0.211(16) X_s  ~,\\
g^\text{loop}_{a\Lambda\Lambda} & = 			-0.095(7) +0.099(9) X_u +0.099(9) X_d -0.149(9) X_s  ~,\\
g^\text{loop}_{a\Sigma^0\Lambda} & = \phantom{-}0.030(3) -0.089(7) X_u +0.089(7) X_d + 0.000(1) X_s ~,
\end{split}
\end{align}
where we have set the scale at $\lambda = 1\,\mathrm{GeV}$. The corresponding results for the
KSVZ and DFSZ models are displayed in Table~\ref{tab:gABloop}. As expected, the loop
contributions are indeed subleading, where the orders of magnitude of the individual
terms range between low $\mathcal{O}\left(10^{-1}\right)$ and $\mathcal{O}\left(10^{-2}\right)$.
 Note that just using $F_\pi$ or $F_p$ for all the decay constants does not lead
to any notable change in these results since the formal difference is of higher order, $\mathcal{O}(p^5)$, in the chiral expansion.

It is remarkable that the loop corrections to
the axion-proton vertex are about one tenth of the full SU(2) result~\cite{Vonk:2020zfh}, which shows that
in this particular case the three-flavor expansion works similar to the case of the
magnetic moments~\cite{Meissner:1997hn}, but different to the baryon masses~\cite{Borasoy:1996bx}
or weak hyperon decays~\cite{Bijnens:1985kj}. Consequently, the largest uncertainty in these
calculations is related to the values of the LECs, as discussed next.
\begin{table}
\begin{tabular}{c|c|c|c|c|c}
\hline
\multirow{3}{*}{Process} & \multicolumn{5}{|c}{$g^\text{loop}_{AB}$} \\ \cline{2-6}
& \multirow{2}{*}{KSVZ} & \multicolumn{4}{|c}{DFSZ} \\ \cline{3-6}
& & general & ${\sin^2\beta=0}$ & ${\sin^2\beta=\tfrac{2}{3}}$ & ${\sin^2\beta=1}$\\
\hline
$\Sigma^+ \to\Sigma^+ + a$ &  $\phantom{-}0.096(7)$ &  $\phantom{-} 0.162(9) - 0.116(7) \sin^2\beta$ &$\phantom{-} 0.162(9)$ &$\phantom{-} 0.085(8)$ & $\phantom{-} 0.046(8)$\\
$\Sigma^- \to\Sigma^- + a$ & $\phantom{-}0.046(6)$ & $\phantom{-} 0.064(9) - 0.019(7) \sin^2\beta$ &$\phantom{-} 0.064(9)$ & $\phantom{-} 0.051(6)$ & $\phantom{-}0.045(8)$\\
$\Sigma^0 \to\Sigma^0 + a$ & $\phantom{-}0.072(7)$ & $\phantom{-} 0.114(9) - 0.068(7) \sin^2\beta$ & $\phantom{-} 0.114(9)$ & $\phantom{-} 0.069(8)$ & $\phantom{-} 0.046(8)$ \\
$p \to p + a$ & 				$\phantom{-}0.046(5)$ & $\phantom{-} 0.100(6) - 0.094(5) \sin^2\beta$ &$\phantom{-} 0.100(6)$ & $\phantom{-} 0.038(6)$ & $\phantom{-} 0.007(6)$\\
$\Xi^- \to\Xi^- + a$ &  				$-0.112(8)$   & 			$- 0.122(11) + 0.038(8) \sin^2\beta$ &	$- 0.122(11)$ & $ -0.096(8)$ & $- 0.083(8)$ \\
$n \to n + a$ & 						$-0.028(5)$ & 			$- 0.046(7) + 0.051(5) \sin^2\beta$ & $- 0.046(7)$ & $- 0.012(6)$ & $\phantom{-} 0.05(6)$ \\
$\Xi^0 \to\Xi^0 + a$ & 				$-0.145(11)$ & 			$- 0.186(13) + 0.102(8) \sin^2\beta$ & $- 0.186(13)$ & $- 0.118(12)$ & $- 0.084(12)$ \\
$\Lambda \to\Lambda + a$ & 			$-0.095(7)$ & 			$- 0.112(8) + 0.050(5) \sin^2\beta$ & $- 0.112(8)$ & $- 0.079(7)$& $- 0.062(8)$ \\ \hline
$\Sigma^0 \to\Lambda + a$ & \multirow{2}{*}{$\phantom{-}0.030(3) $} & \multirow{2}{*}{$\phantom{-} 0.060(4) - 0.059(3) \sin^2\beta$} & \multirow{2}{*}{$\phantom{-} 0.060(4)$} & \multirow{2}{*}{$\phantom{-} 0.020(3)$} & \multirow{2}{*}{$\phantom{-} 0.001(4)$} \\
$\Lambda \to\Sigma^0 + a$ &  &  & & & \\ \hline
\end{tabular}
\caption{One-meson loop contributions to the axion-baryon couplings $g^\text{loop}_{AB}$, Eq~\eqref{eq:NNLOloop}, for the KSVZ axion and the DFSZ axion calculated at scale $\lambda= 1\,\mathrm{GeV}$.}
\label{tab:gABloop}
\end{table}

As for the tree-level contributions from the NNLO Lagrangian (see section~\ref{ch:NNLOchip}), we stated already that the values of the
involved LECs are undetermined hitherto. The scale-dependent parts  $\propto d^r_k(\lambda)$ are expected to compensate
the scale-dependence of the loop contributions, such that the actual observable, $g_{aAB}$, remains 
scale-independent at $\mathcal{O}\left(p^3\right)$. For the following estimation of these hitherto
unknown LECs, we therefore only consider the scale-independent part such that we can leave aside 
the scale-dependent loop contributions in our final estimation of the axion-baryon coupling at
$\mathcal{O}\left(p^3\right)$. Our understanding of the problem is a Bayesian
one: while formally each LEC may take on any arbitrary value, we nevertheless expect that with
sufficient probability the LECs are restricted to values that lead to 
NNLO contributions to $g_{aAB}$ of roughly the same order as the loop contributions discussed above.
In other words: we assume that these contributions are indeed \textit{sub}-leading in comparison
to the leading order contributions, which is basically a naturalness
argument~\cite{Manohar:1983md,Schindler:2009wu}. That this assumption is justified in the present case directly follows from  the numerical results of the loop corrections discussed before.
This argument is of course not universally valid, but it is nevertheless appropriate in a Bayesian
sense, meaning that our results 
from this ansatz can be used as priors in future determinations of the LECs once suitable experimental  or lattice QCD data is available for fitting procedures. Although it is not expected that the axion-nucleon coupling will be measured experimentally with sufficient accuracy in the near future, one may use lattice QCD to compute the relevant couplings by introducing an external isoscalar axial source into the QCD action to mimic the axion,  and compute the corresponding form factors.

In practice, we performed a Monte Carlo sampling of the ten involved LECs 
$d^r_{41}$, \dots, $d^r_{47}$ and $d_{43}^i$, $d_{44}^i$, and $d_{46}^i$ within a reasonable range of $\mathcal{O}\left(1\,\mathrm{GeV}^{-2}\right)$ and extracted those sets of LECs that lead to NNLO corrections to $g_{aAB}$ of low $\mathcal{O}\left(10^{-1}\right)$. In particular, we set $0.15$ as a numerical constraint, which is rather conservative (in view of the the loop contributions given above). The allowed regions for the values of the respective LECs are then given by probability distributions of Gau{\ss}ian type centered around zero (as the overall sign of the NNLO corrections is in principle undetermined from this method). With this approach, one obtains
\begin{align}
d^r_{41} & = 0.00(4)\,\text{GeV}^{-2}, & d^r_{42} & = 0.00(4)\,\text{GeV}^{-2},\nonumber\\
d^r_{43} & = 0.00(4)\,\text{GeV}^{-2}, & d^{i,r}_{43} & = 0.00(11)\,\text{GeV}^{-2},\nonumber\\
d^r_{44} & = 0.00(6)\,\text{GeV}^{-2}, & d^{i,r}_{44} & = 0.00(17)\,\text{GeV}^{-2},\label{eq:LECsBayes1}\\
d^r_{45} & = 0.00(6)\,\text{GeV}^{-2}, & d^r_{46} & = 0.00(11)\,\text{GeV}^{-2}, \nonumber\\
d^{i,r}_{46} & = 0.00(14)\,\text{GeV}^{-2}, & d^r_{47} & = 0.00(14)\,\text{GeV}^{-2}\nonumber.
\end{align}
Moreover, one finds that some of the extrapolated probability distributions of the LECs are correlated. The most important correlation coefficients are given by
\begin{align*}
\operatorname{corr}(d^r_{41},d^r_{44}) & = 0.72~, & \operatorname{corr}(d^r_{41},d^r_{46}) & = -0.59~, \\
\operatorname{corr}(d^r_{41},d^r_{47}) & = -0.84~, & \operatorname{corr}(d^r_{44},d^r_{46}) & = -0.89~, \\
\operatorname{corr}(d^r_{44},d^r_{47}) & = -0.85~, & \operatorname{corr}(d^r_{46},d^r_{47}) & = 0.70~, \\
\operatorname{corr}(d^r_{42},d^r_{45}) & = -0.85~, & \operatorname{corr}(d^{i,r}_{44},d^{i,r}_{46}) & = -0.84~,
\end{align*}
while all other correlation coefficients are negligibly small.

If one additionally considers the large-$N_c$ approach, where $N_c$ is the number of colors, it is
to be expected that the LECs for the terms with two flavor traces ($d_{45,46,47}^r$ and $d_{46}^{i,r}$)
are suppressed relative to those with only one flavor trace  ($d_{41,42,43,44}^r$ and $d_{43,44}^{i,r}$)
by $\mathcal{O}\left(1/N_c\right)$; see, e.g., Refs.~\cite{Gasser:1984gg,Manohar:1998xv}. Therefore,
we performed another Monte Carlo sampling for the LECs, where this expectation is taken into account,
namely by assigning a larger probability to such sets obeying this expected rule in comparison to sets
of LECs deviating from it, which are considered less probable. The result is
\begin{align*}
d^r_{41} & = 0.00(3)\,\text{GeV}^{-2}, & d^r_{42} & = 0.00(2)\,\text{GeV}^{-2},\\
d^r_{43} & = 0.00(4)\,\text{GeV}^{-2}, & d^{i,r}_{43} & = 0.00(10)\,\text{GeV}^{-2},\\
d^r_{44} & = 0.00(3)\,\text{GeV}^{-2}, & d^{i,r}_{44} & = 0.00(13)\,\text{GeV}^{-2},\\
d^r_{45} & = 0.00(1)\,\text{GeV}^{-2}, & d^r_{46} & = 0.00(1)\,\text{GeV}^{-2},\\
d^{i,r}_{46} & = 0.00(7)\,\text{GeV}^{-2}, & d^r_{47} & = 0.00(1)\,\text{GeV}^{-2}.
\end{align*}
As expected, the probability distributions of the rather suppressed LECs become considerably thinner, even though it is not excluded that they in reality may achieve higher values. For the following estimation of the axion-baryon coupling at NNLO, however, we stick to the less rigid estimation of the LECs given in Eq.~\eqref{eq:LECsBayes1}.

The last contribution to the $\mathcal{O}\left(q^3\right)$ axion-baryon coupling stem from terms $\propto \chi_-$ discussed in section~\ref{ch:NNLOchim}. From the matching with SU(2), Eq.~\eqref{eq:xhimcontrSU2}, we deduce that
\begin{equation}
(d_{18}+2 d_{19})_\text{SU(2)} = (d_2+3d_3) \left(1+\frac{w}{1+z}\right)^{-1}~,
\end{equation}
where neither the value of the LEC $d_{19}$ from the SU(2) case, nor the values of $d_2$ and $d_3$ from the SU(3) case are known. The LEC $d_{18}$ 
is fixed by the Goldberger--Treiman discrepancy and given by~\cite{Hoferichter:2015tha}
\begin{equation}
d_{18} = -0.44(24)\,\text{GeV}^{-2}~.
\end{equation}
Using this matching with SU(2) and applying the same Bayesian approach as 
described above, we extrapolate
\begin{equation}
d_{2} = -0.45(24)\,\text{GeV}^{-2}~,\qquad d_{3} = 0.15(92)\,\text{GeV}^{-2}~.
\end{equation}
Finally, we can collect all contributions to estimate the full $\mathcal{O}\left(p^3\right)$ axion-baryon couplings given by
\begin{equation}\label{eq:gaABNNLO}
g_{aAB} = g_{AB} + \frac{ 4 M_{\pi^\pm}^2 z}{1+z}\left(\hat{d}^r_{AB} -\frac{d_2 + 3d_3}{1+z+w} \delta_{AB}\right) -\frac{1}{27} g_{AB}^{\text{loop},r}~,
\end{equation}
which results in
\begin{align}\label{eq:gNNLO}
\begin{split}
g_{a\Sigma^+\Sigma^+} & = -0.547(84)+0.850(98) X_u -0.032(88) X_d -0.455(93) X_s ~,\\
g_{a\Sigma^-\Sigma^-} & = -0.245(80)-0.030(88) X_u +0.852(99) X_d -0.456(93) X_s ~,\\
g_{a\Sigma^0\Sigma^0} & = -0.399(78)+0.420(63) X_u + 0.397(63)X_d -0.456(93) X_s ~,\\
g_{app} & = -0.432(86) + 0.836(99)X_u -0.418(91) X_d -0.053(84) X_s ~,\\
g_{a\Xi^-\Xi^-} & =  \phantom{-}0.166(79)-0.083(84) X_u -0.455(91) X_d +0.854(97) X_s  ~,\\
g_{ann} & =  \phantom{-}0.003(83)-0.398(90) X_u + 0.856(99)X_d -0.053(84) X_s ~,\\
g_{a\Xi^0\Xi^0} & =  \phantom{-}0.303(81)-0.424(91) X_u -0.052(84) X_d +0.854(96) X_s  ~,\\
g_{a\Lambda\Lambda} & =\phantom{-}0.138(87) -0.159(74) X_u -0.137(74) X_d +0.663(92) X_s  ~,\\
g_{a\Sigma^0\Lambda} & = -0.161(24) +0.441(68) X_u -0.497(68) X_d + 0.012(24) X_s ~.
\end{split}
\end{align}
The corresponding results for the KSVZ and the DFSZ axion are collected in Table~\ref{tab:gABaxbarNNLO}. Note that while in the leading order case 
the uncertainties arise from the errors of the quark ratios $z$ and $w$, and the nucleon matrix elements $\Delta u$, $\Delta d$, and $\Delta s$, the uncertainties in the next-to-next-to-leading order case are dominated by the lack of knowledge of the involved LECs.
\begin{table}
\begin{tabular}{c|c|c|c|c|c}
\hline
\multirow{3}{*}{Process} & \multicolumn{5}{|c}{$g_{aAB}$} \\ \cline{2-6}
& \multirow{2}{*}{KSVZ} & \multicolumn{4}{|c}{DFSZ} \\ \cline{3-6}
& & general & ${\sin^2\beta=0}$ & ${\sin^2\beta=\tfrac{2}{3}}$ & ${\sin^2\beta=1}$\\
\hline
$\Sigma^+ \to\Sigma^+ + a$ &  $-0.547(84)$ &  $- 0.709(94) + 0.446(54) \sin^2\beta$ &$- 0.709(94)$ &$- 0.412(88)$ & $- 0.263(91)$\\
$\Sigma^- \to\Sigma^- + a$ & $-0.245(80)$ & $- 0.113(92) - 0.142(54) \sin^2\beta$ &$- 0.113(92)$ & $- 0.208(84)$ & $-0.255(85)$\\
$\Sigma^0 \to\Sigma^0 + a$ & $-0.399(78)$ & $- 0.417(87) + 0.158(43) \sin^2\beta$ & $- 0.417(87)$ & $- 0.311(80)$ & $- 0.259(81)$ \\
$p \to p + a$ & $-0.432(86)$ & $- 0.589(96) + 0.436(53) \sin^2\beta$ &$- 0.589(96)$ & $- 0.298(90)$ & $- 0.153(92)$\\
$\Xi^- \to\Xi^- + a$ &  $\phantom{-}0.166(79)$ & $\phantom{-} 0.299(91) - 
0.161(52) \sin^2\beta$ &$\phantom{-} 0.299(91)$ & $\phantom{-} 0.192(83)$ 
& $\phantom{-} 0.138(84)$ \\
$n \to n + a$ & $\phantom{-}0.003(83)$ & $\phantom{-} 0.271(94) - 0.400(53) \sin^2\beta$ & $\phantom{-} 0.271(94)$ & $\phantom{-} 0.004(87)$ & $- 0.130(88)$ \\
$\Xi^0 \to\Xi^0 + a$ & $\phantom{-}0.303(81)$ & $\phantom{-} 0.570(92) - 0.409(52) \sin^2\beta$ & $\phantom{-} 0.570(92)$ & $\phantom{-} 0.298(85)$ & $\phantom{-} 0.162(87)$ \\
$\Lambda \to\Lambda + a$ & $\phantom{-}0.138(87)$ & $\phantom{-} 0.314(96) - 0.228(47) \sin^2\beta$ & $\phantom{-} 0.314(96)$ & $\phantom{-} 0.161(90)$& $\phantom{-} 0.085(90)$ \\ \hline
$\Sigma^0 \to\Lambda + a$ & \multirow{2}{*}{$-0.161(24) $} & \multirow{2}{*}{$- 0.323(33) + 0.309(32) \sin^2\beta$} & \multirow{2}{*}{$- 0.323(33)$} & \multirow{2}{*}{$- 0.117(30)$} & \multirow{2}{*}{$- 0.014(33)$} \\
$\Lambda \to\Sigma^0 + a$ &  &  & & & \\ \hline
\end{tabular}
\caption{Axion-baryon couplings $g_{aAB}$, Eq~\eqref{eq:gaABNNLO}, for the KSVZ axion and the DFSZ axion at $\mathcal{O}\left(p^3\right)$.}
\label{tab:gABaxbarNNLO}
\end{table}

\section{Summary}\label{sec:summ}

In this paper, we have worked out the axion-baryon coupling in SU(3) HBCHPT up to $\mathcal{O}\left(q^3\right)$
in the chiral power counting and found --- in the case of the axion-nucleon coupling constants --- good
agreement with the already known results obtained in the SU(2) chiral approach. One of the most important
outcomes of this study is that the axion-baryon coupling strengths are all of roughly the same order
for all members of the baryon octet with only few exceptions. The most prominent and well-known example
is the axion-neutron coupling that might vanish in the KSVZ and the DSFZ model for particular values
of $\sin^2\beta$. Given the fact that the axion couples to hyperons with similar strength as it couples to nucleons (or even stronger, especially if the coupling to neutrons is suppressed), our results suggest a revision of axion emissivity of dense stellar objects such as neutron stars, where the cores might contain strange matter in large amounts, as have been proposed in the literature (see the References given in the introduction).

In this study, we considered rather ``traditional'' models. Tree-level axion-quark interactions, if any,
are of the same order for all flavors, and there are no flavor-changing processes. Our calculations, however,
can in principle be extended to flavor non-conserving processes by adjusting the matrix $\mathcal{X}_q$
accordingly and follow the strategy  described in Eq.~\eqref{eq:externalcurrents2}. This then
would lead to new terms in the axion-baryon interaction Lagrangian including baryon-changing processes. 

The other modification of our calculations would be to consider models in 
which the couplings of the axion
to the charm, bottom, and top quark are much stronger than the couplings to the up, down, and strange quarks.
Such a model can easily lead to very strong axion-nucleon couplings driven by sea-quark  effects (thus
avoiding the problem of vanishing axion-neutron coupling), which in the end are balanced by
correspondingly larger values of $f_a$, as has bee shown recently in Ref.~\cite{Darme:2020gyx}. Such ideas are
entirely compatible with our calculations: While the formulae derived in section~\ref{ch:determiningcoupl} would
be unaffected, the only difference would be that the numerical calculations of section~\ref{ch:results}
have to be adjusted as the terms $\propto \Delta q_i$ of \eqref{eq:DFchoice} can not be neglected any more.

Our studies using both SU(2) and SU(3) symmetry have shown that the axion-baryon couplings for rather
standard axion models are now known to good precision, but also that in the next-to-next-to-leading order
case a higher precision is currently unattainable due to the lack of knowledge of some LECs. In
this study, we used a Monte Carlo sampling procedure to extrapolate the most probable values of these
unknown LECs, where ``probable" here has to be understood in a Bayesian sense.  It should thus be
clear that our numerical leading order results, Eq.~\eqref{eq:gLO}, are more solid than the
numerical estimations of the NNLO results.

Once the knowledge of
the parameters in questions is enhanced, the numerical results of this work can easily be updated. However,
the current uncertainties of $g_{aAB}$ are not the major concern considering the fact that the axion window
in terms of $f_a$ is still very large. The largest uncertainties regarding the existence of axions
is hence strongly linked to the uncertainty of $f_a$ and in the end it is 
not $g_{aAB}$ that counts,
but $G_{aAB}=-g_{aAB}/f_a$, see Eqs.~\eqref{eq:generalG1} and \eqref{eq:generalG2}.  At last, we suggest that $G_{aAB}$ may be computed using lattice QCD by introducing an external isoscalar axial source to mimic the axion.

\begin{acknowledgments}
This work is supported in part by the Deutsche Forschungsgemeinschaft (DFG)
and the National Natural Science Foundation of China (NSFC) through the funds provided to the
Sino-German Collaborative Research Center ``Symmetries and the Emergence of Structure in QCD"
(NSFC Grant No. 12070131001, DFG
Project-ID 196253076 -- TRR 110), by the NSFC under Grants No. 11835015 and No. 12047503, by the Chinese Academy of Sciences (CAS) under Grant No. 
QYZDB-SSW-SYS013 and No. XDB34030000,  by the CAS 
President's International Fellowship
Initiative (PIFI) (Grant No.~2018DM0034), by the VolkswagenStiftung (Grant No. 93562), and by the
EU (STRONG2020). 
\end{acknowledgments}

\appendix
\section{SU(3) generators in the physical basis and structure constants}\label{appendix1}

In the present work, we mainly make use of the physical basis~\cite{Krause:1990xc}, based on a set of traceless, non-Hermitian matrices $\lt{A}$, $A=\left\{1,\dots,8\right\}$, such that the baryon octet matrix $B$, Eq.~\eqref{eq:baryons}, and the pseudoscalar meson octet matrix $\Phi$, Eq.~\eqref{eq:mesons}, can be decomposed as
\begin{align}
\begin{split}
B & = \frac{1}{\sqrt{2}} \sum_A \lt{A} B_A~, \\
\bar{B} & = \frac{1}{\sqrt{2}} \sum_A \bar{B}_A \lt{A}^\dagger ~,\\
\Phi & = \sum_{A} \lt{A} \Phi_A~,
\end{split}
\end{align}
where the baryon fields $B_A$ and the meson fields $\Phi_A$ can directly be equated with the physical particles, i.e.
\begin{align}
B_1 & = \Sigma^+~, & B_2 & = \Sigma^-~, & B_3 & = \Sigma^0~, & B_4 & = p~,\nonumber\\
B_5 & = \Xi^-~, & B_6 & = n~, & B_7 & = \Xi^0~, &  B_8 & = \Lambda~. 
\end{align}
and
\begin{align}\label{eq:mesonsexplicit}
\Phi_1 & = \pi^+~, & \Phi_2 & = \pi^-~, & \Phi_3 & = \pi^0~, & \Phi_4 & = K^+~,\nonumber \\
\Phi_5 & = K^-~, & \Phi_6 & = K^0~, & \Phi_7 & = \bar{K}^0~, &  \Phi_8 & = \eta~. 
\end{align}
The explicit form of the generators $\lt{A}$ is
\begin{align}
\begin{split}
\lt{1} & = \lt{2}^\dagger = \begin{pmatrix}
0 & \sqrt{2} & 0 \\
0 & 0 & 0 \\
0 & 0 & 0
\end{pmatrix}~, \quad
\lt{4} =\lt{5}^\dagger = \begin{pmatrix}
0 & 0 & \sqrt{2} \\
0 & 0 & 0 \\
0 & 0 & 0
\end{pmatrix}~, \quad
\lt{6} =\lt{7}^\dagger=\begin{pmatrix}
0 & 0 & 0 \\
0 & 0 & \sqrt{2} \\
0 & 0 & 0
\end{pmatrix}~, \\
\lt{3} & = \begin{pmatrix}
\cos\epsilon+\frac{1}{\sqrt{3}}\sin\epsilon & 0 & 0 \\
0 & -\cos\epsilon+\frac{1}{\sqrt{3}}\sin\epsilon & 0 \\
0 & 0 & -\frac{2}{\sqrt{3}}\sin\epsilon
\end{pmatrix}~, \\
\lt{8} & = \begin{pmatrix}
-\sin\epsilon+\frac{1}{\sqrt{3}}\cos\epsilon & 0 & 0 \\
0 & \sin\epsilon+\frac{1}{\sqrt{3}}\cos\epsilon & 0 \\
0 & 0 & -\frac{2}{\sqrt{3}}\cos\epsilon
\end{pmatrix}~,
\end{split}
\end{align}
where $\epsilon$ is the mixing angle given in Eq.~\eqref{eq:epsmixing}. These matrices are related to the common Gell-Mann matrices $\lambda_a$ by 
a unitary $8\times 8$ transformation matrix $N$ via
\begin{equation}
\lt{a}=\sum_a N_{Aa} \lambda_{a}~, \qquad \lambda_a=\sum_A N^*_{Aa} \lt{A}=\sum_A N_{Aa} \lt{A}^\dagger~.
\end{equation}
The non-zero matrix elements of $N$ are (note the sign convention that deviates from the one in~\cite{Krause:1990xc})
\begin{align}
N_{11} = N_{44} = & N_{66} = \frac{1}{\sqrt{2}}~, & N_{21} = N_{54} = N_{76} & = \frac{1}{\sqrt{2}}~,\nonumber \\
N_{22} = N_{55} = & N_{77} = -\frac{i}{\sqrt{2}}~, & N_{12} = N_{45} = N_{67} & = \frac{i}{\sqrt{2}}~, \\
N_{33} = N_{88} = & \cos{\epsilon}~, & N_{38} = - N_{83} & = \sin{\epsilon} \nonumber
\end{align}
and obey
\begin{equation}
\sum_{a} N_{Aa}^* N_{Ba} = \delta_{AB} , \qquad \sum_A N_{Aa}^* N_{Ab} = \delta_{ab}~.
\end{equation}
As the Gell-Mann matrices, the $\lt{A}$'s are ortho-normalized to a value 
of 2,
\begin{equation}\label{eq:ltnorm}
\tr{\lt{A}^\dagger\lt{B}} = 2 \delta_{AB}~.
\end{equation}
Moreover, they satisfy the commutator and anticommutator relations
\begin{align}\label{eq:commutatorrelations}
\begin{split}
\com{\lt{A}^\dagger}{\lt{B}} & =\sum_C \ft{ABC} \lt{C}^\dagger= \sum_C \ft{BAC} \lt{C}~,\\
\com{\lt{A}}{\lt{B}} & =\sum_C \ft{CAB} \lt{C}~,\\
\com{\lt{A}^\dagger}{\lt{B}^\dagger} &=\sum_C \ft{CBA} \lt{C}^\dagger~,\\
\acom{\lt{A}^\dagger}{\lt{B}} & =\frac{4}{3} \delta_{AB} \mathbbm{1} + \sum_C \dt{ABC} \lt{C}^\dagger =\frac{4}{3} \delta_{AB} \mathbbm{1} + \sum_C \dt{BAC} \lt{C}~,
\end{split}
\end{align}
which gives the product rule
\begin{align}
\begin{split}
\lt{A}^\dagger \lt{B} & =\frac{2}{3}\delta_{AB} \mathbbm{1} + \frac{1}{2}\sum_C \left(\dt{ABC} + \ft{ABC}\right) \lt{C}^\dagger\\ & = \frac{2}{3}\delta_{AB} \mathbbm{1} + \frac{1}{2}\sum_C \left(\dt{BAC} + \ft{BAC}\right) \lt{C}~.
\end{split}
\end{align}
The structure constants defined in Eq.~\eqref{eq:commutatorrelations} can 
be evaluated using
\begin{align}
\begin{split}
\ft{ABC} & = \frac{1}{2}\tr{\lt{A}^\dagger \com{\lt{B}}{\lt{C}}}~,\\
\dt{ABC} & = \frac{1}{2}\tr{\lt{A}^\dagger \acom{\lt{B}}{\lt{C}}}~,
\end{split}
\end{align}
and are related to the corresponding structure constants of the Gell-Mann 
representation via
\begin{align}
\begin{split}
\ft{ABC} & = 2i\sum_{a,b,c} N_{Aa}^* N_{Bb} N_{Cc} f^{abc}~,\\
\dt{ABC} & = 2\sum_{a,b,c} N_{Aa}^* N_{Bb} N_{Cc} d^{abc}~.
\end{split}
\end{align}
In contrast to the structure constants $f^{abc}$ ($d^{abc}$) of the Gell-Mann representation, which are totally anti-symmetric (symmetric), the structure constants $\ft{ABC}$ ($\dt{ABC}$) of the physical basis are anti-symmetric (symmetric) in the last two indices only. Finally, there are sum rules for the structure constants:
\begin{align}\label{eq:sumrules}
\begin{split}
\sum_{C,D} \ft{ACD} \ft{BCD} & = \sum_{C,D} \ft{CAD} \ft{CBD} = 12 \delta_{AB}~,\\
\sum_{C,D} \dt{ACD} \dt{BCD} & = \sum_{C,D} \dt{CAD} \dt{CBD} = \frac{20}{3} \delta_{AB}~.
\end{split}
\end{align}
Often, we also need the generators at $\epsilon=0$, and therefore it is 
convenient to set
\begin{align}\label{eq:lambdahut}
\begin{split}
\lh{A} & =\left. \lt{A}\right|_{\epsilon=0}~,\\
\fh{ABC} & =\left. \ft{ABC}\right|_{\epsilon=0}~,\\
\dh{ABC} & =\left. \dt{ABC}\right|_{\epsilon=0}~.
\end{split}
\end{align}
The properties Eqs.~\eqref{eq:ltnorm}--\eqref{eq:sumrules} are then of course also valid for the $\lh{A}$'s with $\ft{ABC}$ and $\dt{ABC}$ replaced by $\fh{ABC}$ and $\dh{ABC}$, respectively. Note that now one has
\begin{equation}
\lh{3}=\lambda_3~, \qquad \lh{8}=\lambda_8~.
\end{equation}
The non-zero elements of the structure constants in this representation are
\begin{align}
\begin{split}
\fh{147} & = -\fh{256} = \fh{416} =-\fh{527} = \fh{624} = -\fh{715} = \sqrt{2}~,\\
\fh{443} & = -\fh{553} = -\fh{663} = \fh{773} = -\fh{345} =\fh{367}= -1~,\\
\fh{448} & = -\fh{558} = \fh{668} = -\fh{778} = -\fh{845} =-\fh{867}= -\sqrt{3}~,\\
\fh{113} & = -\fh{223} = -\fh{312} = -2~.
\end{split}
\end{align}
Similarly, 
\begin{align}
\begin{split}
\dh{147} & = \dh{256} = \dh{416} = \dh{527} = \dh{624} = \dh{715} = \sqrt{2}~,\\
\dh{443} & = \dh{553} = -\dh{663} = -\dh{773} = \dh{345} = -\dh{367}= 1~,\\
\dh{118} & = \dh{228} = \dh{338} = \dh{812} = \dh{833} =-\dh{888}= \frac{2}{\sqrt{3}}~,\\
\dh{448} & = \dh{558} = \dh{668} = \dh{778} = \dh{845} = \dh{867} = -\frac{1}{\sqrt{3}}~.
\end{split}
\end{align}

\section{Matrix elements of \texorpdfstring{{\boldmath $g_{AB}$, $\hat{d}_{AB}$, and $g_{AB}^{\text{loop}}$}}{gAB, dAB and gABloop}}\label{appendix2}
Here we collect the matrix elements of the several contributions to the axion-baryon coupling. Table~\ref{tab:gAB} shows the matrix elements of $g_{AB}$ as defined in Eq.~\eqref{eq:gAB} in the physical basis at $\epsilon\neq 0$, see Eq.~\eqref{eq:epsmixing}.
\begin{table}[hbt]
\begin{tabular}{c|c|c}
\hline
Process & $A=B$ &  $g_{AB}$ \\
\hline
$\Sigma^+ \to\Sigma^+ + a$ & 1 &  $2\left(\frac{c^{(8)}}{\sqrt{3}} D + c^{(3)}F\right)+ c_i D^i$\\
$\Sigma^- \to\Sigma^- + a$ & 2 & $2\left(\frac{c^{(8)}}{\sqrt{3}} D - c^{(3)}F\right)+ c_i D^i$\\
$\Sigma^0 \to\Sigma^0 + a$ & 3 & $ \phantom{-}\frac{2}{\sqrt{3}}\left( c^{(3)}\sin 2\epsilon + c^{(8)} \cos 2\epsilon\right)D+ c_i D^i$\\
$p \to p + a$ & 4 & $\phantom{-}c^{(3)} (D+ F)- \frac{c^{(8)}}{\sqrt{3}} (D-3F) + c_i D^i$\\
$\Xi^- \to\Xi^- + a$ & 5 & $\phantom{-}c^{(3)} (D- F)- \frac{c^{(8)}}{\sqrt{3}} (D+3F) + c_i D^i$\\
$n \to n + a$ & 6 & $-c^{(3)} (D+ F)- \frac{c^{(8)}}{\sqrt{3}} (D-3F)  + c_i D^i$\\
$\Xi^0 \to\Xi^0 + a$ & 7 & $-c^{(3)} (D- F)- \frac{c^{(8)}}{\sqrt{3}} (D+3F) + c_i D^i$\\
$\Lambda \to\Lambda + a$ & 8 & $- \frac{2}{\sqrt{3}}\left( c^{(3)}\sin 2\epsilon + c^{(8)} \cos 2\epsilon\right)D+ c_i D^i$\\
\hline
Process & $A\neq B$ &  $g_{AB}$ \\
\hline
$\Sigma^0 \to\Lambda + a$ & $A=8,~B=3$ & \multirow{2}{*}{$\frac{2}{\sqrt{3}}\left( c^{(3)}\cos 2\epsilon + c^{(8)} \sin 2\epsilon\right)D $}\\
$\Lambda \to\Sigma^0 + a$ & $A=3,~B=8$ & \\ \hline
\end{tabular}
\caption{Non-zero matrix elements of $g_{AB}$, Eq.~\eqref{eq:gAB}.}
\label{tab:gAB}
\end{table}

In Eqs.~\eqref{eq:d11}--\eqref{eq:d83}, we list the non-zero matrix elements of the next-to-next-to-leading order contributions to the axion-baryon coupling as defined in Eqs.~\eqref{eq:xhipcontr} and \eqref{eq:dABhut}. 
Here we refrain from explicitly marking the scale dependence of the LECs $d^{(i)}_k(\lambda)$.
\begin{align}\label{eq:d11}
\begin{split}
\hat{d}_{11} = & ~  c^{(3)} \biggl\{ \left(4d_{41}+d_{47}\right)\left(1-\frac{1}{z}\right) + 4 d_{42} \left(1+\frac{1}{z}\right)+ 2d_{45}\left(1+\frac{1}{z}+\frac{1}{w}\right) \biggr\}\\
& + \frac{c^{(8)}}{\sqrt{3}} \biggl\{ 4d_{43}\left(1-\frac{1}{z}\right)  + 4 d_{44} \left(1+\frac{1}{z}\right)+ 2d_{46}\left(1+\frac{1}{z}+\frac{1}{w}\right)\\ & \qquad\qquad + d_{47}\left(1+\frac{1}{z}-\frac{2}{w}\right) \biggr\}\\
& + c_i \biggl\{ d_{43}^i\left(1-\frac{1}{z}\right)  + d_{44}^i \left(1+\frac{1}{z}\right)+ d_{46}^i\left(1+\frac{1}{z}+\frac{1}{w}\right)\biggr\}
\end{split}
\end{align}
\begin{align}\label{eq:d22}
\begin{split}
\hat{d}_{22} = & ~  c^{(3)} \biggl\{ \left(4d_{41}+d_{47}\right)\left(1-\frac{1}{z}\right)  - 4 d_{42} \left(1+\frac{1}{z}\right)-2 d_{45}\left(1+\frac{1}{z}+\frac{1}{w}\right)\biggr\}\\
& +\frac{c^{(8)}}{\sqrt{3}} \biggl\{-4d_{43}\left(1-\frac{1}{z}\right)  + 
4 d_{44} \left(1+\frac{1}{z}\right)+ 2d_{46}\left(1+\frac{1}{z}+\frac{1}{w}\right) \\
& \qquad\qquad + d_{47}\left(1+\frac{1}{z}-\frac{2}{w}\right)
\biggr\}\\
& + c_i \biggl\{ -d_{43}^i\left(1-\frac{1}{z}\right)  + d_{44}^i \left(1+\frac{1}{z}\right)+ d_{46}^i\left(1+\frac{1}{z}+\frac{1}{w}\right)\biggr\}
\end{split}
\end{align}
\begin{align}\label{eq:d33}
\begin{split}
\hat{d}_{33} = & ~  c^{(3)} \left(4d_{44}+d_{47}\right)\left(1-\frac{1}{z}\right)  \\
& +\frac{c^{(8)}}{\sqrt{3}} \biggl\{4 d_{44} \left(1+\frac{1}{z}\right)+ 2d_{46}\left(1+\frac{1}{z}+\frac{1}{w}\right) + d_{47}\left(1+\frac{1}{z}-\frac{2}{w}\right) \biggr\}\\
& + c_i \biggl\{ d_{44}^i \left(1+\frac{1}{z}\right)+ d_{46}^i\left(1+\frac{1}{z}+\frac{1}{w}\right)\biggr\}
\end{split}
\end{align}
\begin{align}\label{eq:d44}
\begin{split}
\hat{d}_{44} = & ~ c^{(3)} \biggl\{ 2\left( d_{41}+d_{43} \right)\left(1-\frac{1}{w}\right)+2\left(d_{42}+d_{44} \right)\left(1+\frac{1}{w}\right) \\
& \qquad +\left(d_{45}+d_{46} \right)\left(1+\frac{1}{z}+\frac{1}{w}\right) +d_{47} \left(1-\frac{1}{z}\right)
\biggr\}\\
& +\frac{c^{(8)}}{\sqrt{3}} \biggl\{ 2\left(3 d_{41}-d_{43} \right)\left(1-\frac{1}{w}\right)+2\left(3d_{42}-d_{44} \right)\left(1+\frac{1}{w}\right) \\
& \qquad + \left(3d_{45}-d_{46} \right) \left(1+\frac{1}{z}+\frac{1}{w}\right) +d_{47} \left(1+\frac{1}{z}-\frac{2}{w}\right)
\biggr\}\\
& + c_i \biggl\{ d_{43}^i\left(1-\frac{1}{w}\right) + d_{44}^i \left(1+\frac{1}{w}\right)+ d_{46}^i\left(1+\frac{1}{z}+\frac{1}{w}\right)\biggr\}
\end{split}
\end{align}
\begin{align}\label{eq:d55}
\begin{split}
\hat{d}_{55} = & ~ c^{(3)} \biggl\{2\left( d_{41}-d_{43} \right)\left(1-\frac{1}{w}\right)-2\left(d_{42}-d_{44} \right)\left(1+\frac{1}{w}\right) \\
& \qquad -\left(d_{45}-d_{46} \right)\left(1+\frac{1}{z}+\frac{1}{w}\right) +d_{47} \left(1-\frac{1}{z}\right)
\biggr\}\\
& +\frac{c^{(8)}}{\sqrt{3}} \biggl\{ 2\left( 3d_{41}+d_{43} \right)\left(1-\frac{1}{w}\right)-2\left(3d_{42}+d_{44} \right)\left(1+\frac{1}{w}\right)\\
& \qquad - \left(3d_{45}+d_{46} \right) \left(1+\frac{1}{z}+\frac{1}{w}\right) +d_{47} \left(1+\frac{1}{z}-\frac{2}{w}\right)
\biggr\}\\
& + c_i \biggl\{ -d_{43}^i\left(1-\frac{1}{w}\right) + d_{44}^i \left(1+\frac{1}{w}\right)+ d_{46}^i\left(1+\frac{1}{z}+\frac{1}{w}\right)\biggr\}
\end{split}
\end{align}
\begin{align}\label{eq:d66}
\begin{split}
\hat{d}_{66} = & ~ c^{(3)} \biggl\{-2\left( d_{41}+d_{43} \right)\left(\frac{1}{z}-\frac{1}{w}\right)-2\left(d_{42}+d_{44} \right)\left(\frac{1}{z}+\frac{1}{w}\right) \\
& \qquad -\left(d_{45}+d_{46} \right)\left(1+\frac{1}{z}+\frac{1}{w}\right) +d_{47} \left(1-\frac{1}{z}\right)
\biggr\}\\
& +\frac{c^{(8)}}{\sqrt{3}} \biggl\{2\left(3 d_{41}-d_{43} \right)\left(\frac{1}{z}-\frac{1}{w}\right)+2\left(3d_{42}-d_{44} \right)\left(\frac{1}{z}+\frac{1}{w}\right) \\
& \qquad + \left(3d_{45}-d_{46} \right)\left(1+\frac{1}{z}+\frac{1}{w}\right) +d_{47} \left(1+\frac{1}{z}-\frac{2}{w}\right)
\biggr\}\\
& + c_i \biggl\{ d_{43}^i\left(\frac{1}{z}-\frac{1}{w}\right) + d_{44}^i \left(\frac{1}{z}+\frac{1}{w}\right)+ d_{46}^i\left(1+\frac{1}{z}+\frac{1}{w}\right)\biggr\}
\end{split}
\end{align}
\begin{align}\label{eq:d77}
\begin{split}
\hat{d}_{77} = & ~ c^{(3)} \biggl\{-2\left( d_{41}-d_{43} \right)\left(\frac{1}{z}-\frac{1}{w}\right)+2\left(d_{42}-d_{44} \right)\left(\frac{1}{z}+\frac{1}{w}\right) \\
& \qquad +\left(d_{45}-d_{46} \right)\left(1+\frac{1}{z}+\frac{1}{w}\right) +d_{47} \left(1-\frac{1}{z}\right)
\biggr\}\\
& +\frac{c^{(8)}}{\sqrt{3}} \biggl\{-2\left(3 d_{41}+d_{43} \right)\left(\frac{1}{z}-\frac{1}{w}\right)-2\left(3d_{42}+d_{44} \right)\left(\frac{1}{z}+\frac{1}{w}\right) \\
& \qquad - \left(3d_{45}+d_{46} \right)\left(1+\frac{1}{z}+\frac{1}{w}\right) +d_{47} \left(1+\frac{1}{z}-\frac{2}{w}\right)
\biggr\}\\
& + c_i \biggl\{ -d_{43}^i\left(\frac{1}{z}-\frac{1}{w}\right) + d_{44}^i 
\left(\frac{1}{z}+\frac{1}{w}\right)+ d_{46}^i\left(1+\frac{1}{z}+\frac{1}{w}\right)\biggr\}
\end{split}
\end{align}
\begin{align}\label{eq:d88}
\begin{split}
\hat{d}_{88} = & ~ c^{(3)} \left(\frac{4}{3}d_{44}+d_{47}\right)\left(1-\frac{1}{z}\right)\\
& +\frac{c^{(8)}}{\sqrt{3}} \biggl\{ \frac{4}{3}d_{44} \left(1+\frac{1}{z}-\frac{8}{w}\right)-2d_{46}\left(1+\frac{1}{z}+\frac{1}{w}\right) + d_{47}\left(1+\frac{1}{z}-\frac{2}{w}\right)
\biggr\}\\
& + c_i \biggl\{\frac{1}{3} d_{44}^i \left(1+\frac{1}{z}+\frac{4}{w}\right)+ d_{46}^i\left(1+\frac{1}{z}+\frac{1}{w}\right)\biggr\}
\end{split}
\end{align}
For the only non-diagonal matrix elements, one has $\hat{d}_{38}=\hat{d}_{83}$, which is given by
\begin{equation}\label{eq:d83}
\hat{d}_{38} = \frac{c^{(3)}}{\sqrt{3}} \biggl\{ 4d_{44} \left(1+\frac{1}{z}\right)+d_{46}\left(1+\frac{1}{z}+\frac{1}{w}\right) \biggr\} +  \left( \frac{4c^{(8)}}{3} d_{44} + \frac{c_i}{\sqrt{3}} d_{44}^i\right) \left(1-\frac{1}{z}\right)
\end{equation}
Eqs.~\eqref{eq:gloop11}--\eqref{eq:gloop38} show the matrix elements of the finite one meson loop contributions $g_{AB}^{\text{loop},r}$, cf. Eq.~\eqref{eq:gABloopr}, of diagram (a) (Fig.~\ref{fig:oneloop}). The expressions have been simplified using the leading order relations~\eqref{eq:mesonmasses} and \eqref{eq:mixingmesonmasses}. Moreover, we have set $\mu_{\Phi_C} = M_{\Phi_C}/(4\pi F_p)$ for any meson $\Phi_C$.
\begin{align}\label{eq:gloop11}
\begin{split}
g_{11}^{\text{loop},r} = & ~  c^{(3)} \biggl\{ -3D\left(3D^2-11F^2\right)\Mkmk \\ & \qquad - F \left(D^2-9F^2\right)\left(4 \mu^2_{\pi^\pm}+\mu^2_{K^\pm} +\mu^2_{K^0}\right) -24D^2 F \mu_{\pi^\pm}^2 \biggr\}\\
& + \frac{c^{(8)}}{\sqrt{3}} \biggl\{ -D\left(D^2+9F^2\right)\left(16 \mu^2_{\pi^\pm}+\mu^2_{K^\pm} +\mu^2_{K^0}\right) \\ & \qquad +54DF^2\left(4 
\mu^2_{\pi^\pm}-\mu^2_{K^\pm} -\mu^2_{K^0}\right)-3F\left(7D^2+9F^2\right)\Mkmk \biggr\}\\
& + c_i D^i \biggl\{30DF\Mkmk + \left(D^2+9F^2\right)\left(4 \mu^2_{\pi^\pm}+\mu^2_{K^\pm} +\mu^2_{K^0}\right) \\ & \qquad +12D^2\left(\mu^2_{K^\pm} +\mu^2_{K^0}\right) \biggr\}
\end{split}
\end{align}
\begin{align}\label{eq:gloop22}
\begin{split}
g_{22}^{\text{loop},r} = & ~  c^{(3)} \biggl\{ -3D\left(3D^2-11F^2\right)\Mkmk \\ & \qquad + F \left(D^2-9F^2\right)\left(4 \mu^2_{\pi^\pm}+\mu^2_{K^\pm} +\mu^2_{K^0}\right) +24D^2 F \mu_{\pi^\pm}^2 \biggr\}\\
& + \frac{c^{(8)}}{\sqrt{3}} \biggl\{ -D\left(D^2+9F^2\right)\left(16 \mu^2_{\pi^\pm}+\mu^2_{K^\pm} +\mu^2_{K^0}\right) \\ & \qquad +54DF^2\left(4 
\mu^2_{\pi^\pm}-\mu^2_{K^\pm} -\mu^2_{K^0}\right)+3F\left(7D^2+9F^2\right)\Mkmk \biggr\}\\
& + c_i D^i \biggl\{-30DF\Mkmk + \left(D^2+9F^2\right)\left(4 \mu^2_{\pi^\pm}+\mu^2_{K^\pm} +\mu^2_{K^0}\right) \\ & \qquad +12D^2\left(\mu^2_{K^\pm} +\mu^2_{K^0}\right) \biggr\}
\end{split}
\end{align}
\begin{align}\label{eq:gloop33}
\begin{split}
g_{33}^{\text{loop},r} = & ~  c^{(3)} D\left(17D^2-9F^2\right)\Mkmk\\
& + \frac{c^{(8)}}{\sqrt{3}} \biggl\{-D\left(D^2+9F^2\right)\left(16 \mu^2_{\pi^\pm}+\mu^2_{K^\pm} +\mu^2_{K^0}\right) \\ & \qquad +54DF^2\left(4 \mu^2_{\pi^\pm}-\mu^2_{K^\pm} -\mu^2_{K^0}\right) \biggr\}\\
& + c_i D^i \biggl\{ \left(D^2+9F^2\right)\left(4 \mu^2_{\pi^\pm}+\mu^2_{K^\pm} +\mu^2_{K^0}\right) +12D^2\left(\mu^2_{K^\pm} +\mu^2_{K^0}\right)\biggr\}
\end{split}
\end{align}
\begin{align}\label{eq:gloop44}
\begin{split}
g_{44}^{\text{loop},r} = & ~  c^{(3)} \biggl\{-D\left(D^2+3F^2\right) \left(5 \mu^2_{\pi^\pm}+8\mu^2_{K^\pm} -4\mu^2_{K^0}\right) + 60DF^2\Mkmk\\ &\qquad -D^2F \left(11 \mu^2_{\pi^\pm}+14\mu^2_{K^\pm} -10\mu^2_{K^0}\right) - 9F^3\left( \mu^2_{\pi^\pm}-2\mu^2_{K^\pm} -2\mu^2_{K^0}\right) \biggr\}\\
& + \frac{c^{(8)}}{\sqrt{3}} \biggl\{-D^3\left(13 \mu^2_{\pi^\pm}-8\mu^2_{K^\pm} -14\mu^2_{K^0}\right)+9DF^2\left(9 \mu^2_{\pi^\pm}-4\mu^2_{K^\pm} 
-2\mu^2_{K^0}\right)\\ & \qquad +3D^2F\left(3 \mu^2_{\pi^\pm}-14\mu^2_{K^\pm} -4\mu^2_{K^0}\right) + 27F^3\left(\mu^2_{\pi^\pm}+2\mu^2_{K^\pm}\right) \biggr\}\\
& + c_i D^i \biggl\{30DF\left( \mu^2_{\pi^\pm} -\mu^2_{K^0}\right) +\left(D^2+9F^2\right)\left(\mu^2_{\pi^\pm}+4\mu^2_{K^\pm} +\mu^2_{K^0}\right) \\ & \qquad +12D^2\left( \mu^2_{\pi^\pm} +\mu^2_{K^0}\right) \biggr\}
\end{split}
\end{align}
\begin{align}\label{eq:gloop55}
\begin{split}
g_{55}^{\text{loop},r} = & ~  c^{(3)} \biggl\{-D\left(D^2+3F^2\right) \left(5 \mu^2_{\pi^\pm}+8\mu^2_{K^\pm} -4\mu^2_{K^0}\right) + 60DF^2\Mkmk\\ &\qquad +D^2F \left(11 \mu^2_{\pi^\pm}+14\mu^2_{K^\pm} -10\mu^2_{K^0}\right) + 9F^3\left( \mu^2_{\pi^\pm}-2\mu^2_{K^\pm} -2\mu^2_{K^0}\right) \biggr\}\\
& + \frac{c^{(8)}}{\sqrt{3}} \biggl\{-D^3\left(13 \mu^2_{\pi^\pm}-8\mu^2_{K^\pm} -14\mu^2_{K^0}\right)+9DF^2\left(9 \mu^2_{\pi^\pm}-4\mu^2_{K^\pm} 
-2\mu^2_{K^0}\right)\\ & \qquad -3D^2F\left(3 \mu^2_{\pi^\pm}-14\mu^2_{K^\pm} -4\mu^2_{K^0}\right) - 27F^3\left(\mu^2_{\pi^\pm}+2\mu^2_{K^\pm}\right) \biggr\}\\
& + c_i D^i \biggl\{-30DF\left( \mu^2_{\pi^\pm} -\mu^2_{K^0}\right) +\left(D^2+9F^2\right)\left(\mu^2_{\pi^\pm}+4\mu^2_{K^\pm} +\mu^2_{K^0}\right) 
\\ & \qquad +12D^2\left( \mu^2_{\pi^\pm} +\mu^2_{K^0}\right) \biggr\}
\end{split}
\end{align}
\begin{align}\label{eq:gloop66}
\begin{split}
g_{66}^{\text{loop},r} = & ~  c^{(3)} \biggl\{D\left(D^2+3F^2\right) \left(5 \mu^2_{\pi^\pm}-4\mu^2_{K^\pm} +8\mu^2_{K^0}\right) + 60DF^2\Mkmk\\ 
&\qquad +D^2F \left(11 \mu^2_{\pi^\pm}-10\mu^2_{K^\pm} +14\mu^2_{K^0}\right) + 9F^3\left( \mu^2_{\pi^\pm}-2\mu^2_{K^\pm} -2\mu^2_{K^0}\right) \biggr\}\\
& + \frac{c^{(8)}}{\sqrt{3}} \biggl\{-D^3\left(13 \mu^2_{\pi^\pm}-14\mu^2_{K^\pm} -8\mu^2_{K^0}\right)+9DF^2\left(9 \mu^2_{\pi^\pm}-2\mu^2_{K^\pm} 
-4\mu^2_{K^0}\right)\\ & \qquad +3D^2F\left(3 \mu^2_{\pi^\pm}-4\mu^2_{K^\pm} -14\mu^2_{K^0}\right) + 27F^3\left(\mu^2_{\pi^\pm}+2\mu^2_{K^0}\right) \biggr\}\\
& + c_i D^i \biggl\{30DF\left( \mu^2_{\pi^\pm} -\mu^2_{K^\pm}\right) +\left(D^2+9F^2\right)\left(\mu^2_{\pi^\pm}+\mu^2_{K^\pm} +4\mu^2_{K^0}\right) \\ & \qquad +12D^2\left( \mu^2_{\pi^\pm} +\mu^2_{K^\pm}\right) \biggr\}
\end{split}
\end{align}
\begin{align}\label{eq:gloop77}
\begin{split}
g_{77}^{\text{loop},r} = & ~  c^{(3)} \biggl\{D\left(D^2+3F^2\right) \left(5 \mu^2_{\pi^\pm}-4\mu^2_{K^\pm} +8\mu^2_{K^0}\right) + 60DF^2\Mkmk\\ 
&\qquad -D^2F \left(11 \mu^2_{\pi^\pm}-10\mu^2_{K^\pm} +14\mu^2_{K^0}\right) - 9F^3\left( \mu^2_{\pi^\pm}-2\mu^2_{K^\pm} -2\mu^2_{K^0}\right) \biggr\}\\
& + \frac{c^{(8)}}{\sqrt{3}} \biggl\{-D^3\left(13 \mu^2_{\pi^\pm}-14\mu^2_{K^\pm} -8\mu^2_{K^0}\right)+9DF^2\left(9 \mu^2_{\pi^\pm}-2\mu^2_{K^\pm} 
-4\mu^2_{K^0}\right)\\ & \qquad -3D^2F\left(3 \mu^2_{\pi^\pm}-4\mu^2_{K^\pm} -14\mu^2_{K^0}\right) - 27F^3\left(\mu^2_{\pi^\pm}+2\mu^2_{K^0}\right) \biggr\}\\
& + c_i D^i \biggl\{-30DF\left( \mu^2_{\pi^\pm} -\mu^2_{K^\pm}\right) +\left(D^2+9F^2\right)\left(\mu^2_{\pi^\pm}+\mu^2_{K^\pm} +4\mu^2_{K^0}\right) \\ & \qquad +12D^2\left( \mu^2_{\pi^\pm} +\mu^2_{K^\pm}\right) \biggr\}
\end{split}
\end{align}
\begin{align}\label{eq:gloop88}
\begin{split}
g_{88}^{\text{loop},r} = & ~ -5 c^{(3)} D\left(D^2-9F^2\right)\Mkmk \\
& + \frac{c^{(8)}}{\sqrt{3}} \biggl\{D\left(D^2+27F^2\right)\left(\mu^2_{K^\pm} +\mu^2_{K^0}\right) +4D^3\left(10 \mu^2_{\pi^\pm}-3\mu^2_{K^\pm} -3\mu^2_{K^0}\right)\biggr\}\\
& + c_i D^i \biggl\{3\left(D^2+9F^2\right)\left(\mu^2_{K^\pm} +\mu^2_{K^0}\right)-4D^2\left(4 \mu^2_{\pi^\pm}+\mu^2_{K^\pm} +\mu^2_{K^0}\right) \biggr\}
\end{split}
\end{align}
The only non-diagonal matrix elements are given by
\begin{align}\label{eq:gloop38}
\begin{split}
g_{38}^{\text{loop},r} = & ~  \frac{c^{(3)}}{\sqrt{3}}D \biggl\{\left(D^2-9F^2\right) \left(8 \mu^2_{\pi^\pm}-\mu^2_{K^\pm} -\mu^2_{K^0}\right) + 8D^2 \left(\mu^2_{\pi^\pm}-2\mu^2_{K^\pm} -2\mu^2_{K^0}\right) \biggr\}\\
& + \left\{ c^{(8)} D \left(11D^2-27F^2\right)  -3\sqrt{3} c_i D^i \left(D^2-3F^2\right)\right\} \Mkmk
\end{split}
\end{align}
and $g_{83}^{\text{loop},r}=g_{38}^{\text{loop},r}$.

\end{document}